\documentclass[aps,prx,twocolumn,superscriptaddress,nofootinbib,longbibliography]{revtex4-2}

\usepackage[utf8]{inputenc}
\usepackage[T1]{fontenc}
\usepackage{lmodern}
\usepackage{amsmath,amssymb,amsfonts}
\usepackage{graphicx}
\usepackage{hyperref}
\usepackage[dvipsnames]{xcolor}
\usepackage{physics}
\usepackage{bm}
\usepackage{braket}
\usepackage{mathtools}
\usepackage{float}
\usepackage{comment}
\usepackage{tikz}
\usepackage{enumitem}
\usepackage{lipsum}
\usepackage{subcaption}
\usepackage{siunitx}
\usepackage{placeins}

\hypersetup{
    colorlinks=true,
    linkcolor=blue,
    citecolor=blue,
    urlcolor=blue
}
\begin{document}
\title{Time-Efficient Quantum Many-Body State Synthesis and its Optimization via Warm Start Strategies}
\author{Prashasti Tiwari}
\email{ucappti@ucl.ac.uk}
\affiliation{Department of Physics, University College London, London WC1E 6BT, United Kingdom}

\author{Dylan Lewis}
\affiliation{Department of Physics, University College London, London WC1E 6BT, United Kingdom}
\affiliation{Blackett Laboratory, Imperial College London, London SW7 2AZ, United Kingdom}

\author{Sougato Bose}
\affiliation{Department of Physics, University College London, London WC1E 6BT, United Kingdom}

\date{\today}

\begin{abstract}
Quantum mechanical ground states of many-body systems can be important resources for various investigations: for quantum sensing, for benchmarking quantum hardware with classically solvable states, as the initial states for nonequilibrium quantum dynamics following quenches, the simulation of quantum processes that start by coupling systems in ground states, eg, could be a process in quantum chemistry, while their approximations are required as inputs to quantum phase estimation algorithm. However, preparing ground states can be challenging; for example, it may require adiabatic switching of Hamiltonian terms slower than an inverse gap, which can be time consuming and bring in decoherence. Here we 
investigate the possibility of preparing a many-body entangled ground state of a certain Hamiltonian, which can be called a quantum ``problem'' Hamiltonian, using the time evolution of an initial fiducial state by another time independent ``solver'' Hamiltonian with couplings up to unit strength for a very short fixed (unit) time: a ``time efficient'' ansatz. The parameters of the solver Hamiltonian are optimised classically minimising energy as the cost function. We present a study of up to $n=14$ qubit many-body states prepared using this methodology. Importantly, we find that a strategy of combining a warm start (an already prepared ground state of a $n-1$ qubit Hamiltonian) and incrementally adding extra couplings of a qubit is the best scaling strategy to prepare the ground state of a $n$-qubit Hamiltonian. 
\end{abstract}

\maketitle

\section{Introduction}


Quantum many-body systems are comprised of a large number of interacting quantum particles, which are typically electrons, spins, or atoms. In such systems, the components exhibit strongly correlated behaviour -- typically, they have entangled states \cite{sachdev2011quantum}. These states are fundamental to understanding physical phenomena in quantum chemistry, nuclear physics, quantum information, phase transitions, topological order, and condensed matter physics. It is important to prepare these states in a controlled manner in a quantum simulator or quantum computer so that the effects of various interactions can be probed \cite{feynman1982simulating,preskill2018quantum}.  Among the varied states of quantum many-body systems it is often pivotal to be able to prepare ground states as these states play a decisive role in their low-energy behaviour. The generation of special ground states is required for quantum error correction codes, spin liquids, and topological quantum states for quantum computing \cite{sachdev2011quantum, vojta2003quantum, greiner2002quantum,li2025low,bespalova2021quantum, Volya2023, bausch2022fast}. Further, the preparation of known quantum states in a quantum hardware can help us to benchmark it. Moreover, one of the really difficult class of problems that only quantum simulators/computers can solve is the long time nonequilibrium dynamics of quantum many-body systems, which underlies several natural processes (true, there are often relaxations in open environments, but one may want to study that without approximations using the closed system dynamics of a system-environment many-body system). Such dynamics typically involves a large part of an exponentially large Hilbert space (for example, $2^{n}$ dimensional for $n$ spin-1/2 particles/qubits) which makes their classical simulation computationally difficult. Ground states of quantum many-body systems would be the necessary prerequisite as initial states to model various such natural processes in quantum simulators: for example, the non-adiabatic dynamics \cite{macdonell2021analog} that ensues after bringing two ground state complex molecules to suddenly interact with each other, or suddenly bringing two many-body ground states to interact at a point \cite{bayat2010entanglement} or how a system changes due to an external change of environment or internal couplings \cite{barmettler2010quantum}, say, by a sudden change of pressure on a solid. They may also be useful for studying thermalization/non-thermalization in closed and open systems following quenches with entangled ground states as starting states \cite{sels2022bath}. Those states that can be prepared may also serve as building blocks for larger \cite{patkowski2026hierarchical} or higher dimensional ground states in a simulator. In fact, the controlled preparation of many-body states has also become a building block for quantum machine learning applications, in hybrid quantum-classical algorithms, and is an important focus for both digital and analog quantum computing platforms  \cite{ward2009preparation,arguello2019analogue,gibbs2025exploiting,umer2025probing,jaderberg2025variational,jaderberg2026learning,singh2026ground}. Last, but not the least, we should emphasize one of the most important applications of quantum state preparation, namely as the input to quantum phase estimation algorithm \cite{kitaev1995quantum,cleve1998quantum}. The quantum phase estimation algorithm can be run to find eigenvalues such as ground state energies to desired precision, but to have any real quantum advantage, one needs an initial quantum state with a high overlap with the ground state \cite{abrams1999quantum,aspuruguzik2005simulated,babbush2018lowdepth,pelofske2026vqe}.

In a nutshell, it is crucial to be able to prepare the quantum ground state of a system in a given quantum hardware, as this can be the starting point of various subsequent investigations. However, it is a challenging problem as even knowing the ground state analytically does not necessarily mean that one can easily translate that to a method of preparing it `in situ' in a quantum simulator or via a quantum circuit.


\subsection*{Current Approaches}
There are many approaches to quantum state preparation. These methods include adiabatic methods \cite{Jansen2007, Guo2024, Rai2024, schaller2008adiabatic, Wan2022}, digital based variational methods, analog quantum simulation \cite{Peruzzo2014, Preti2022, Wang2023, Grimsley2024, Watanabe2024, Shen2022, Gokhale2021} and hybrid analog methods \cite{Geier2023, Crescimanna2023, Geier2025, Bhargava2026}. The adiabatic and analog methods start with an initial state of a Hamiltonian and gradually evolved towards a target state, whilst also remaining in the ground state throughout. However adiabatic methods require that the time scale inversely with spectral energy gaps \cite{roland2002quantum,lubasch2011adiabatic,farooq2015adiabatic}. The analog methods are better suited to trapped-ion and Rydberg atom simulators, but it can become harder to control the decoherence and noise due to the large timescales. There are other techniques proposed such as shortcuts to adiabacity and counteradiabatic driving to overcome the large timescales, however they propose challenges as the quantum fluctuations cause transitions into unwanted final energy states \cite{Calzetta_2018}. There are classical methods such as Matrix Peoduct States (MPS) for finding out the entangled ground states for certain Hamiltonians. They 
represent states with area law entanglement entropy efficiently \cite{Eisert2010, Cramer2006, Refael2009, Stephan2011, Malz2024, Marti_2025, Coopmans2024, Schuckert2024, Scarpa2020, Wei2024, Jahromi2021, Verstraete2004, Nakagawa2020}. However, when converting a MPS representation to a circuit to create the ground state, one naturally has to convert the tensors to a multi-qubit gate so that multiple ancillary qubits are required depending on bond dimension -- in other words, more qubits than those making the ground state are necessary. Moreover, this then naturally require a sequence of gate evolutions acting on the ancillary qubits.

\
On the other hand, digital variational quantum algorithms approach \cite{bharti2022noisy,tilly2022variational, McClean2016} requires a parametrised circuit comprised of unitaries with the energy of the system used as a cost function for the classical optimiser which interacts with the circuit via a feedback loop. In this case, the choice and optimisation of the ansatz remains crucial for the accuracy. The circuit typically contains blocks with some CNOT gates. However, to get to a ground state from some fiducial state, several complete CNOTs may actually not be necessary, eg partial evolutions according to entangling Hamiltonians may suffice. That may take much less time than several full CNOTs (note that some approaches \cite{wecker2015progress,hadfield2019quantum,kokail2019self,lyu2020accelerated,feulner2022variational,grimsley2023adaptive} do not have this issue). These methods become increasingly difficult with the number of parameters in the circuit ansatz, eg the number of layers \cite{larocca2025barren}. It suggests that there is a need of a problem inspired structure for faster convergence and higher fidelity. There has been further work to improve the convergence and fidelity as the system scales, however the scaling problem is not fully eliminated \cite{larocca2025barren,B_rligea_2025,korpas2025undecidableproblemsassociatedvariational,cerezo2021variational, peruzzo2014variational, nature_computational_science_2022, arute2020hartree}. Other methods use VQE through engineering the state\cite{schuld2021effect, wu2025enhancing}. Beyond the hardware-efficient ansatz, there is a large body of work which explore systematic and data driven optimal circuit design, using qubit coupled cluster, adaptive schemes, machine learning, symmetry constraints and genetic algorithms. \cite{ryabinkin2018qubit, kunitski2019double, PhysRevResearch.2.023074, du2022quantum, PRXQuantum.2.010324, kuo2021quantum, altares2021automatic, huang2022robust, sun2024quantum, lyu2023symmetry, han2024multilevel}



\subsection*{Our Approach}
In this work, we ask a fundamental question: can the time evolution, say for a fixed short time ($t=1$, say) by a realistic (two-body) constant Hamiltonian (or a very few, eg, a couple of constant Hamiltonians in succession) enable us to generate the desired ground state of a many-body system from a given fiducial state? The independent parameters of the Hamiltonian ($\sim O(n^2)$) are only polynomial in the number $n$ of qubits. Moreover, by restricting to one invocation of a constant Hamiltonian it is equivalent to a shallow circuit -- just a single multiqubit gate enacted in unit time by the Hamiltonian. While this would seem limited, the exponentiation of the Hamiltonian is equivalent to the exponentiation of the sum of several non-commuting terms: this still gives the ansatz a large enough reach in state space (the nested commutators have the capability to generate a large Lie Algebra, although we must say that the coefficients of the terms are not all independent, restricted by being composed from the $O(n^2)$ independent parameters of the Hamiltonian). Does this approach work well as an {\em analog ansatz} for state preparation?

 To examine the above question, we propose to find the required constant Hamiltonian for time evolution via a variational approach. To have a general/expressible enough (so that it is capable of reaching enough states), yet relatively simple, class of Hamiltonians to work with, we restrict the time evolution Hamiltonian to anisotropic Heisenberg models on arbitrary graphs/chains with arbitrary local fields. Given recent results in trapped ion systems \cite{teoh2019machinelearningdesigntrappedion}, these are on the verge of realisability in a quantum simulator. The parameters of the analog ansatz (the time evolution Hamiltonian) are subjected to a gradient based optimisation. The cost function is the energy of the evolved state with respect to a different Hamiltonian, namely the Hamiltonian whose ground state we want to prepare. It is thus a local cost function -- sum of various local terms. There is the recent body of literature on an obstacle to gradient based optimization, namely exponentially vanishing gradients during the optimization, called the barren plateaus problem \cite{larocca2025barren}. Among the strategies on mitigating barren plateaus in variational circuits: Problem-inspired ansatz choice, initialisation strategies, adaptive iterative methods, alternative cost functions, and noise aware training\cite{cunningham2025investigating}, we naturally adapt the first one. There is also work on a mathematical framework for understanding the case for barren plateaus using Lie algebraic theory \cite{ragone2024lie}, and work on the interplay of both the trainability and expressability in parametrised quantum circuits \cite{chen2024taming}. However, the gradient based minimisation of the cost function is anticipated to work effectively in our context, and is, in fact, expected from a recent proof \cite{park2024hamiltonian} that when the parameters of a quantum circuit are so constrained that they are expressible as the evolution due to a single time independent Hamiltonian, then they do not suffer from the barren plateaus problem. In simple words, the circuit is so shallow with a single multiqubit gate defined by $O(n^2)$ parameters that the problem of exponentially vanishing gradients should not arise.   
 
 Additionally, we incroporate warm start strategies \cite{grimsley2023adaptive,rudolph2023synergistic,dborin2022matrix,mele2022avoiding,puig2025variational, moreno2025generative,lechner2024warm, zhang2024warm} (and combinations of strategies are proposed and examined): a solution for a smaller system is used as a starting point for a bigger system with incremental coupling ramps. We find that this reduces iteration counts and mitigates the gradient disappearing before the solution is reached. Our ansatz and method are benchmarked for a Heisenberg spin graph of up to 6 qubits and a Heisenberg spin-chain for upto 14 qubits to show that high fidelty preparation of ground states of these systems is possible with low enough optimisation times.

\section{Models \& Ansatz}

The systems explored in this work are the spin - 1/2 Heisenberg models. They model a wide variety of systems in nature so that it is important to prepare their ground states for improving our understanding of low energy properties, as well as a starting point of various nonequilibrium dynamics. On the other hand, we will use the same class of models, namely the anisotropic XYZ Heisenberg Hamiltonians, as the analog ansatz to create the required ground states via nonequilibrium dynamics. Such an ansatz is also motivated by the fact that they naturally arise for the ion trap and neutral atom systems, thus implementation can potentially be made by using these systems as quantum simulator platforms. There has been work done to show they arise from the via the laser control of parameters in ion traps \cite{teoh2019machinelearningdesigntrappedion}.

 The Hamiltonian for a $n$-spin anisotropic Heisenberg Hamiltonian system (which we will call our {\em problem} Hamiltonian $H_p$) is as follows
\begin{equation}
    H_p(\mathbf{J}_p)=\sum_{\langle i,j \rangle \in E(G) } \big{(}J^x_{ij}\sigma_i^x\sigma_j^x+J^y_{ij}\sigma_i^y\sigma_j^y+J^z_{ij}\sigma_i^z\sigma_j^z\big{)}
    \label{HXYZ}
\end{equation}
The above is the general equation for the XYZ model, $J^x_{ij}$, $J^y_{ij}$, $J^z_{ij}$ represent the couplings between qubits $i$ and $j$ along each of the spin directions, $E(G)$ is the set of edges for the interaction graph $G$. The aim is to consider a very general model (arbitrary graphs) so that solutions can then be then be used for a broader range of problem sets.

The problem Hamiltonian $H_p$, given by Eq.(\ref{HXYZ}), is used to define the system of interest, whose ground state is to be prepared. This is the very reason that we have decided to call it the problem Hamiltonian. For analogy, note that several classical combinatorial optimization problems can be mapped on to classical Ising Hamiltonians -- the ground state is the solution to the problem. Analogously, $H_p$ presents a {\em quantum problem} (as it has noncommuting terms) to be solved. The quantum (entangled) ground state can be regarded as the solution to the problem. While in general a local Hamiltonian problem is one of the hardest quantum problems -- QMA-complete \cite{kempe2006complexity}, the above $H_p$ may not necessarily be so. In this paper, as problems, two graph topologies are considered i) the chain graph for upto 14 qubits corresponding to a nearest neighbour couplings on a 1D lattice ii) the complex graph for upto 6 qubits to represent all-to-all connectivity. 

 In order to solve the quantum problem (prepare the ground state), we use time evolution by a constant Hamiltonian as discussed above. We call this the {\em solver} Hamiltonian $H_s$. In order to make it slightly more versatile -- as it has to be optimised to solve the problem by time evolution starting from simple fiducial states, $H_s$ is chosen to be the same type of Hamiltonian, but with additional external fields, as given below

\begin{align}
        H_s(\mathbf{J}_s) =\sum_{\langle i,j\rangle  \in E(G) } \big{(}J^x_{ij}\sigma_i^x\sigma_j^x+J^y_{ij}\sigma_i^y\sigma_j^y+J^z_{ij}\sigma_i^z\sigma_j^z\big{)} \nonumber \\ + 
    \sum_{i} \big{(} h_i^x \sigma^x_i + h_i^y \sigma^y_i + h_i^z \sigma^z_i\big{)}
    \label{H_t}
\end{align}
The $h_i$ fields are different for each qubit and present in the $x,y,z$ directions. The above ansatz can be motivated by the realisations of the relevant spin couplings in arbitrary graphs \cite{teoh2019machinelearningdesigntrappedion} for trapped ion qubits, in addition to the fact the local fields can be obtained by appropriately detuned Rabi drivings (however, note that we are taking a bit of academic freedom here in proposing the above $H_s$ as our ansatz, to see how much that empowers us -- the exact mechanism for realisation in a simulator is not part of this work). The time evolution by
$H_s$ is applied to an initial fiducial state of the collection of $n$ qubits, which we take to be either $\ket{0}^{\otimes n}$ or $\ket{+}^{\otimes n}$. 

The final state obtained is therefore given by:

\begin{equation}
    \ket{\psi(\mathbf{J}_s)} = e^{-iH_s(\mathbf{J}_s)t}\ket{\psi_0}
    \label{variationalansatz}
\end{equation}
where $\ket{\psi_0}=\ket{0}^{\otimes N}$ or $\ket{+}^{\otimes N}$.
Moreover, we constrain the values of  $|J_{ij}| \leq 1$, so that the time $t=1$ is a fair estimate of the short duration of time evolution to obtain the ground state, i.e, we do not simply enhance $|J_{ij}$ to decrease the time. Thus we can call our ansatz given by Eqs.(\ref{H_t}) and (\ref{variationalansatz}) with $|J_{ij}| \leq 1$ and $t=1$ as a $\textit{time-efficient}$ ansatz.

Although our ansatz can be generalised to $L$ layers:

\begin{equation}
    \ket{\psi(\{\mathbf{J}_{s,k}\})} = \prod_{k=1}^Le^{-iH_s(\mathbf{J}_{s,k})t}\ket{\psi_0},
    \label{variationalansatzgeneral}
\end{equation}
we will restrict here to $L=1$ (equivalent to very shallow circuits).
Although superficially Eq.(\ref{variationalansatzgeneral}) looks similar to a Hamiltonian variational ansatz it is different. The form of Hamiltonian variational ansatz commonly used evolves according to the separate terms comprising the Hamiltonians $H_s(\mathbf{J}_{s,k})$ or $H_p(\mathbf{J}_{p})$, such as, eg, $J^x_{ij}\sigma_i^x\sigma_j^x$ and has many layers of successive evolution with different Hamiltonian terms. Often these are the terms of $H_p(\mathbf{J}_{p})$ itself \cite{wecker2015progress,grimsley2023adaptive,hadfield2019quantum,lyu2020accelerated}. Perhaps one of the closest in approach is Ref.\cite{kokail2019self}, although there a decay of couplings with distance in an ion trap simulator is assumed, and several layers with interspersed local unitaries are used.

\section{Methods}
Our variational ansatz can be optimized in a classical computer for small sizes or a quantum-classical hybrid approach of a quantum simulator and classical computer. In either way, for each ground state, an optimal solver Hamiltonians $H_s$ would be identified. In some loose sense of the term, then, we do not worry so much about the time it takes to find out the optimal couplings; once we get the $H_s$ corresponding to a given ground state it can be catalogued and repeatedly used in various analog simulators to rapidly generate that ground state. This can then be used as a starting point of further investigations, such as nonequilibrium dynamics. Having said that, it is always best to minimize the time needed to variationally find the optimal $H_s$. To this end, we formulate and compare a few strategies. The cost function for the optimisation is given by 

\begin{align}
    C(\mathbf{J}_s) = \bra{\psi(\mathbf{J}_s)}H_p(\bf{J}_p)\ket{\psi(\mathbf{J}_s)}.
    \label{cost}
\end{align}
As we will only be using classical simulations here to demonstrate that our methodology works,  we will {\em apriori} calculate both the exact ground state $\ket{GS(H_p(\bf{J}_p))}$ and the ground state energy $E_{GS}$ of the problem Hamiltonian $H_p(\bf{J}_p)$ by exact numerical diagonalization. This will help us to keep track of the progress of our optimization (how well we are doing as we progress), although neither will be used in the optimisation algorithm -- only $C(\mathbf{J}_s)$ of Eq.(\ref{cost}) will be used as the relevant cost function. In passing here we note that even in the limited scenario of small sizes where we can numerically exactly compute $\ket{GS(H_p(\bf{J}_p))}$ and $E_{GS}$, our algorithm is still useful as it provides, in the end, a prescription $e^{-iH_s(\mathbf{J}_s)t}\ket{\psi_0}$ to rapidly generate $\ket{GS(H_p(\bf{J}_p))}$ in a quantum simulator.

An iteration is defined as the single optimisation step of the variational optimiser, where parameters are updated once. We also want to keep track of how our optimisation is progressing as we go on iterating. To this end, in addition to the cost function, we compute a few other metrics at each step of iteration.    
 For each iteration, JAX based optimisation is used which also calculates the gradient of the cost function from Eq.(\ref{cost}). This is used to obtain the grad norm $\Vert \nabla C\Vert$, which in turn is used to compute the metric $\beta$
\begin{equation}
    \beta = \frac{\Delta E}{\Vert\nabla C\Vert}
\end{equation}
where $\Delta E$ is the difference in the final energy achieved or the final cost $\bar{C}$ at the end of the latest iteration and the ground state energy known exactly $E_{GS}$: 
\begin{equation}
\Delta E = \bar{C} - E_{\mathrm{GS}}
\label{eq:deltaE}
\end{equation}
The quantity $\beta$ shows how the gradient decays relative to the energy difference.
Another figure of merit computed after each step of iteration is the fidelity $F$ between the variational state 
$\ket{\psi(\mathbf{J}_s)}$ and the exact ground state 
$\ket{GS(H_p(\bf{J}_p))}$ is defined as
\begin{equation}
F(\mathbf{J}_s)
= \left\lvert
    \left\langle GS(H_p(\bf{J}_p)) \mid \psi(\mathbf{J}_s) \right\rangle
  \right\rvert^{2}
  \label{eq:Fidelity}
\end{equation}
 Note that the exact ground state $\ket{GS(H_p(\bf{J}_p))}$ of $H_p(\bf{J}_p)$ had been acquired through numerical exact diagonalisation as stated before, and recorded, which allows us to compute $F(\mathbf{J}_s)$ at each step. The diagnostics allow us to understand the convergence behaviour, where the grad norm $\Vert\nabla C\Vert$ indicates the steepness of the landscape, $\beta$ is a quantitive value for the efficiency of the gradient, and $F(\mathbf{J}_s)$ to understand the accuracy of state synthesis. 

\begin{figure*}[t]
    \centering

    \begin{subfigure}{0.44\textwidth}
        \centering
        \includegraphics[width=\textwidth]{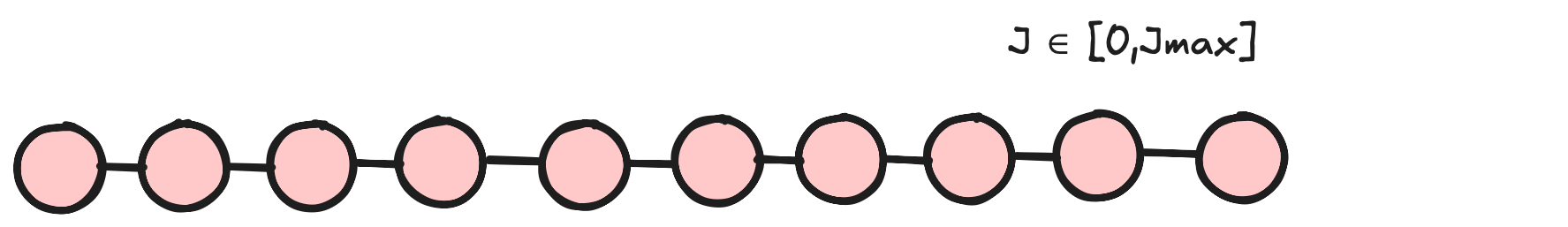}
        \caption{Cold Start}
        \label{fig:ColdStart}
    \end{subfigure}
    \vspace{0.3cm}

    \begin{subfigure}{0.44\textwidth}
        \centering
        \includegraphics[width=\textwidth]{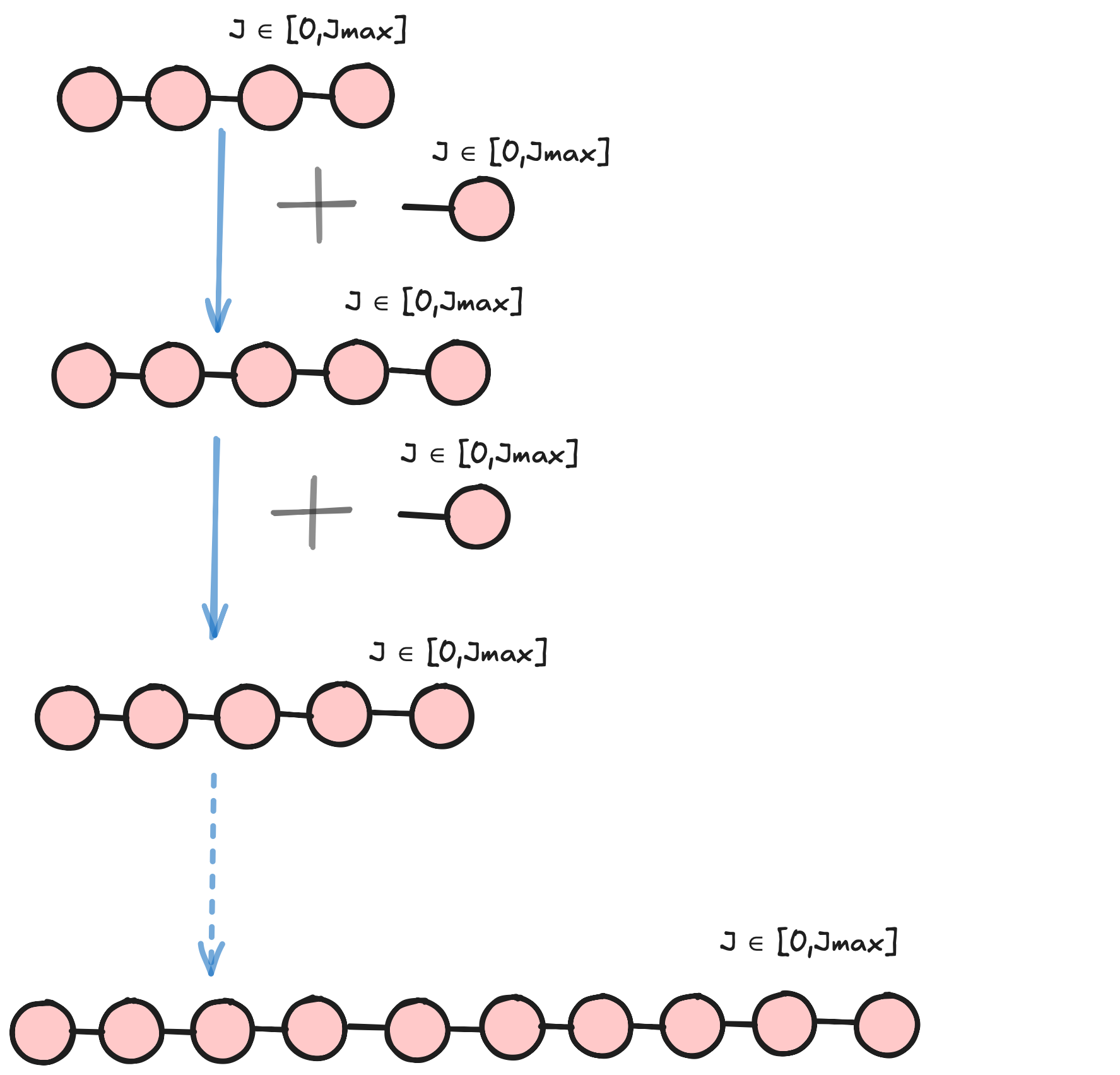}
        \caption{Warm Start by size}
        \label{fig:sub2}
    \end{subfigure}
    \hfill
    \begin{subfigure}{0.44\textwidth}
        \centering
        \includegraphics[width=\textwidth]{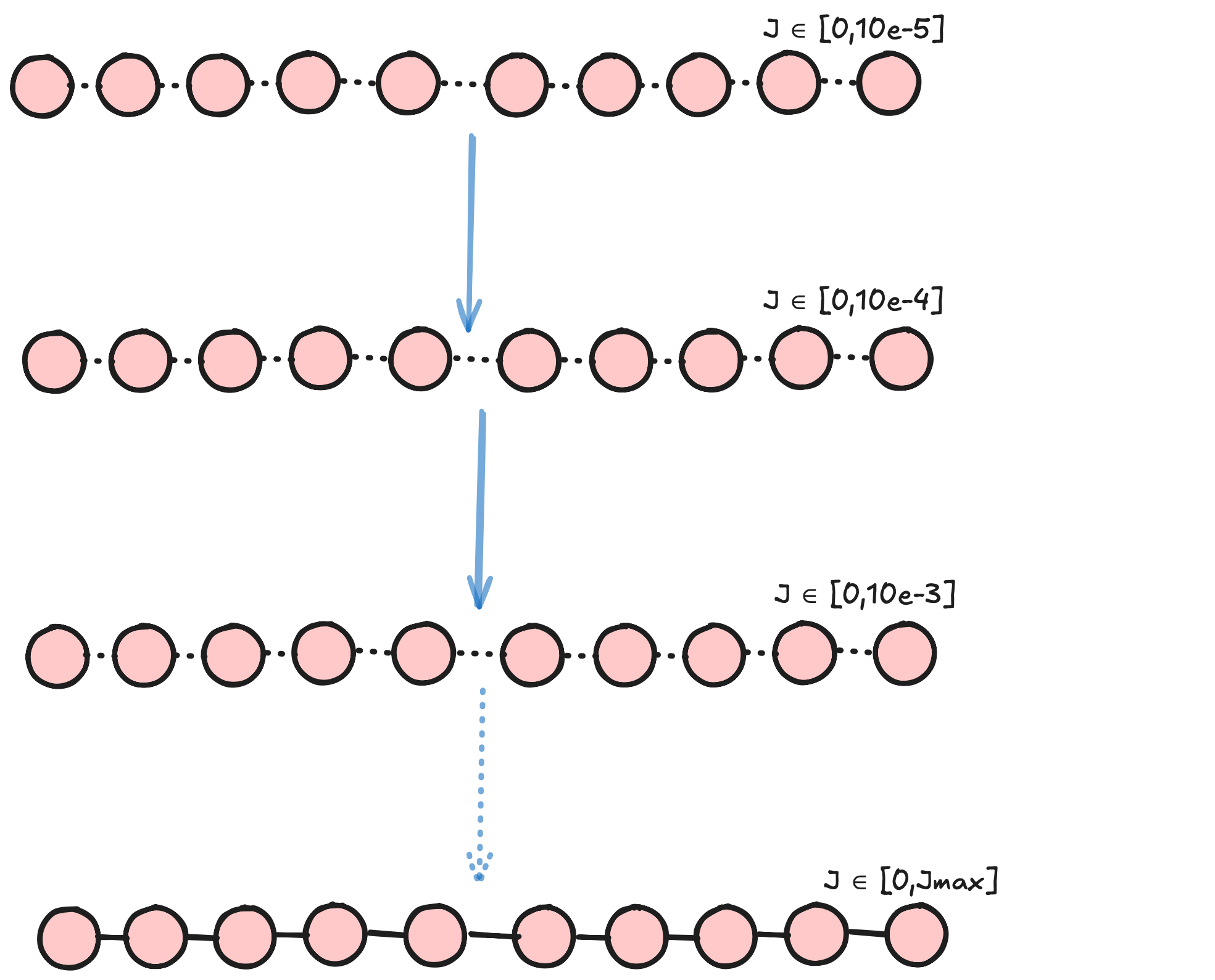}
        \caption{Warm Start Incremental}
        \label{fig:sub3}
    \end{subfigure}
    \vspace{0.3cm}

    \begin{subfigure}{0.99\textwidth}
        \centering
        \includegraphics[width=\textwidth]{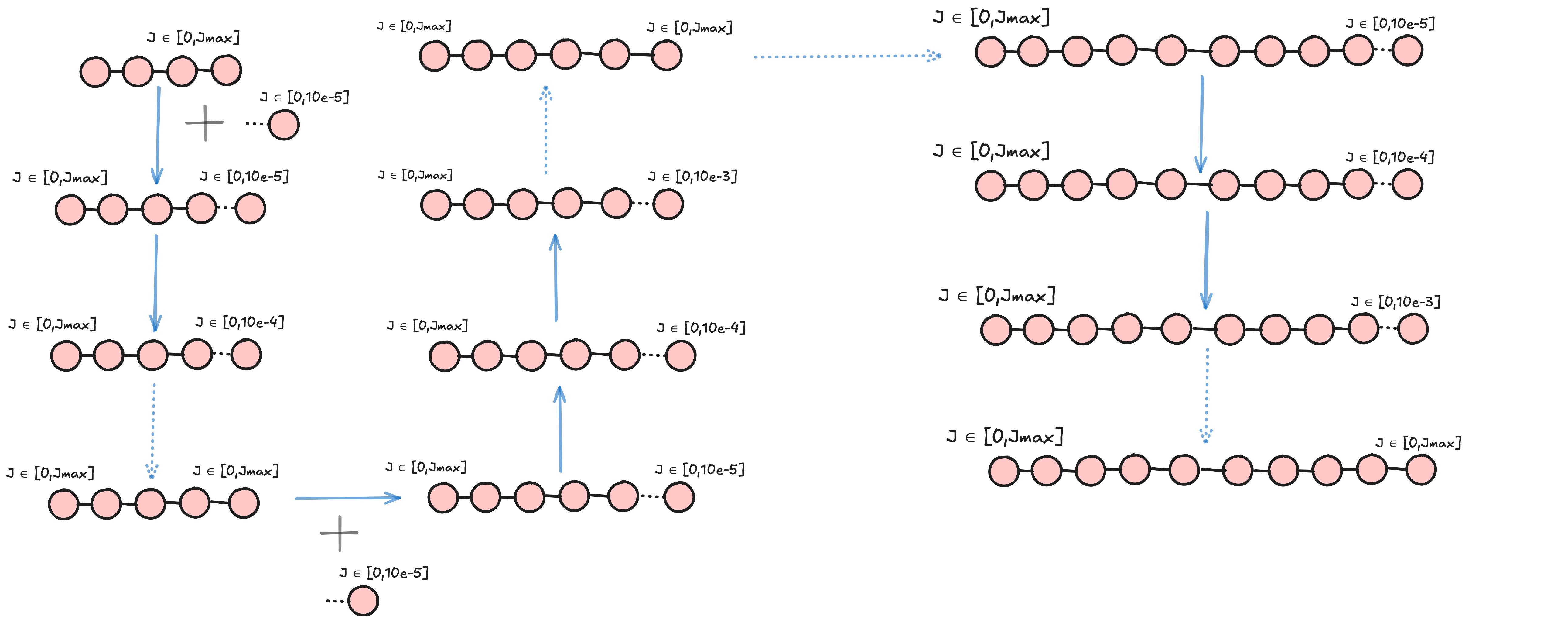}
        \caption{Warm Start Combination}
        \label{fig:sub4}
    \end{subfigure}

    \caption{Illustration of the proposed initialisation strategies for a 10-qubit chain. Chains of size up to 14 qubits are considered, as well as graphs of up to 6 qubits.}
    \label{fig:methodsillustrations}
\end{figure*}

We use certain types of warm start techniques in this study to speed up the convergence to solution. A warm start refers to the use of the knowledge from a related previous optimisation to accelerate the convergence \cite{grimsley2023adaptive,rudolph2023synergistic,dborin2022matrix,mele2022avoiding,puig2025variational}. The four strategies, are illustrated in Fig.\ref{fig:methodsillustrations}; although the figure shows only chains, it is also applied to graphs. The strategies  are:\\

{\bf 1. Warm start by size expansion:} Here the parameters optimised for a $n$-qubit system, then system size grows by adding additional qubits, where the previous parameters of $n$ qubit system form the starting point for the $n+1$ qubit system.
This is shown in Fig.\ref{fig:methodsillustrations}(b).\\ 

{\bf 2. Incremental coupling ramp:} Here the optimization begins with weak strength with $J<<1$ (which is just the shorthand for {\em all} $J^{\alpha}_{ij}<<1$) of all couplings and gradually increases to full strength $J_{\text{max}}\sim 1$ in steps, where the parameters obtained for the previous strength are used for the next round of iterations.This is shown in Fig.\ref{fig:methodsillustrations}(c).\\

{\bf 3. Combination Ramp:} Here both the techniques 1 and 2 are combined, where the system size is increased, then starting with very low coupling strength with the new qubit, the coupling strength is increased till it reaches a maximum $J_{\text{max}}\sim 1$, where the number of qubits is increased. This is shown in Fig.\ref{fig:methodsillustrations}(d).\\

{\bf 4. Cold Start:} Here the optimisation starts for the final number of qubits $n$, with initialisation $\ket{0}^{\otimes n}$ or $\ket{+}^{\otimes n}$
and there are no stored parameters. This is kept as a control method to be able to compare to the others. This is shown in Fig.\ref{fig:methodsillustrations}(a).\\

All the Warm start schemes (1, 2 and 3) use a previous solution of a state that is close to the final state, thus explot the smoothness of the energy landscape, and prevent the gradient from decaying fast, compared to the Cold start, with the aim of obtaining a high fidelty.


\section{Results}

The performance of the above methods and algorithm was studied for both chain and complete-graphs upto $n=14$ and $n=6$ qubits respectively. The simulations used, as sample random problems, $H_p(\bf{J}_p)$ with normalised couplings randomly sampled from a uniform probability distribution with $J_{\text{max}}\sim 1$. Similarly, couplings in the final optimized $H_s(\bf{J}_s)$ are also restricted to a maximum $J_{\text{max}}\sim 1$ so that saying that the evolution for a time $t=1$ is rapid makes sense -- without such a restriction, one can always make stronger couplings for faster dynamics.

The first step of this simulation is to construct a sample problem Hamiltonian $H_p$. The Hamiltonians prepared are either for a chain, i.e. nearest-neighbour interactions (for $n$ up to 14), and for complete graphs (for $n$ up to 6), i.e, long range all-to-all couplings without any particular distance dependent fall offs. As stated above, the couplings are chosen to be of the form given by Eq.(\ref{HXYZ}) and they are sampled from a uniform distribution with maximum $J_{\text{max}}\sim 1$. Now, importantly, for  $H_s$ we choose the {\em same} type of graph, i.e., the non-zero elements are in the same edges of the graph as $H_p$ (whose ground state is to be found) but with {\em additional} external fields (their form indicated in Eq.(\ref{H_t})). We do this just to restrict our optimization over a lower set couplings (and we do find that this is sufficient for success) and with the heuristic that these sites have to be correlated by the dynamics. This is, in effect, a problem inspired ansatz. The external fields are to enable us to flip qubits and go outside the restricted space afforded by only the coupling terms. During the optimization, based on the four methods we have outlined in the previous section, the couplings are sampled from a uniform distribution with a varying maximum $J$.

As we have stated before, instead of a series of gates (a digital parametric quantum circuit) or the common form of a Hamiltonian variational ansatz (a series of evolutions by smaller components of a many-qubit Hamiltonian), the continuous time evolution by a single $H_s$ given by Eq.(\ref{variationalansatzgeneral}) for a time $t=1$ is considered as our single multi-qubit gate involving {\em all} the qubits ($n$-qubit gate). The circuit form of this is illustrated in Fig.\ref{fig:circuit1}. This can be considered as a single layer, $L=1$, variational algorithm. For each choice of $H_s(\bf{J}_s)$, time evolution gives the final quantum evolved quantum state $\ket{\psi(\mathbf{J}_s)}$. 




\subsection{Abitrary Chains}
\begin{figure}[htbp!]
    \centering
    \label{fig:circuit1}{
        \includegraphics[width=0.95\linewidth]{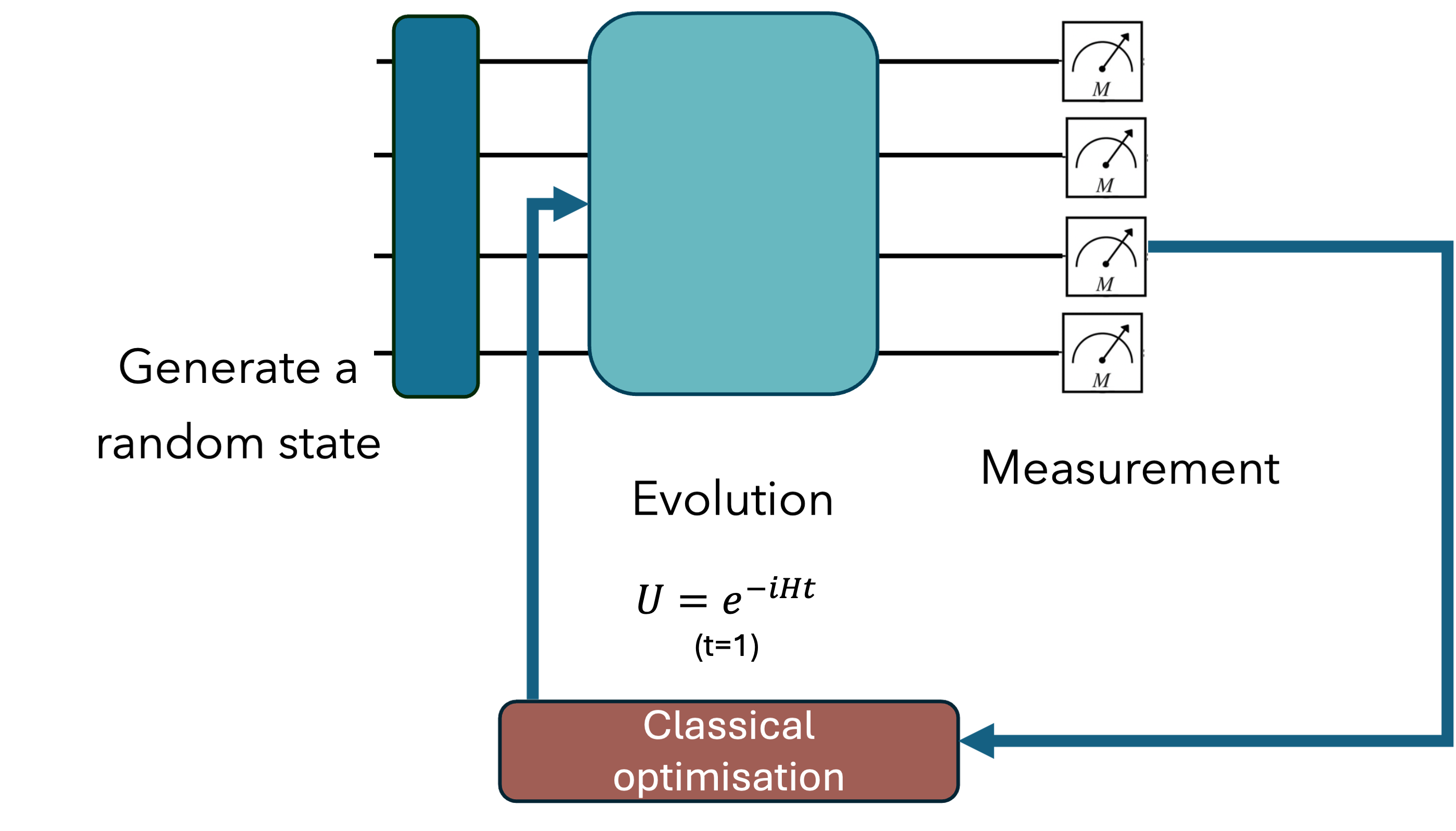}
    }\\[1em]


    \caption{Analog variational optimisation circuits used in this work. One-layer analog variational optimisation circuit ($L=1$) corresponding to the ansatz in Eq.~(\ref{variationalansatz}). The input state is $\ket{+}^{\otimes n}$ or $\ket{0}^{\otimes n}$. A classical optimiser (L-BFGS in this work) updates the parameters.}
    \label{fig:circuit1}
\end{figure}

A JAX-based variational analog algorithm is used. The cost function is the energy expectation value as given in Eq.(\ref{cost}). The process is analogous to variational quantum algorithms run on quantum hardware -- however, here we simulate every step classically and, having full information of $\ket{\psi(\mathbf{J}_s)}$ at each step, do not simulate quantum measurements on the output of the circuit; simply use $\ket{\psi(\mathbf{J}_s)}$ to evaluate the energy cost function. Thus, while the algorithm depicted in Figs.\ref{fig:circuit1}, which shows measurements at the end, is actually the algorithm one would use if one were to implement our method on a quantum simulator for large $n$, here, up to $n=14$ we simply use the $\ket{\psi(\mathbf{J}_s)}$ found after each iteration in our classical simulation.
The optimisation is performed with the use of the L-BFGS algorithm, a quasi-Newton optimisation algorithm, until the convergence was achieved. 

After each iteration step, the quantity $\Delta E$ is lowered. However, the quality of the optimisation after an iteration is characterized by the Fidelity of the evolved state with the true ground state of $H_p$ obtained through diagonalisation. The progress of the optimization is also benchmarked through the metric $\beta$. However, the values of Energy, Fidelity and $\beta$ for a particular sample problem $H_p$ is not indicative of the average behaviour; thus we obtain the above metrics averaged over 100 instances of problem Hamiltonians.

\begin{figure}[htbp!]
    \centering

    \subfloat[Energy difference with respect to iterations at $n=14$, $J = J_{\max}$.
    \label{fig:10chains_a}]{
        \includegraphics[width=0.95\columnwidth]{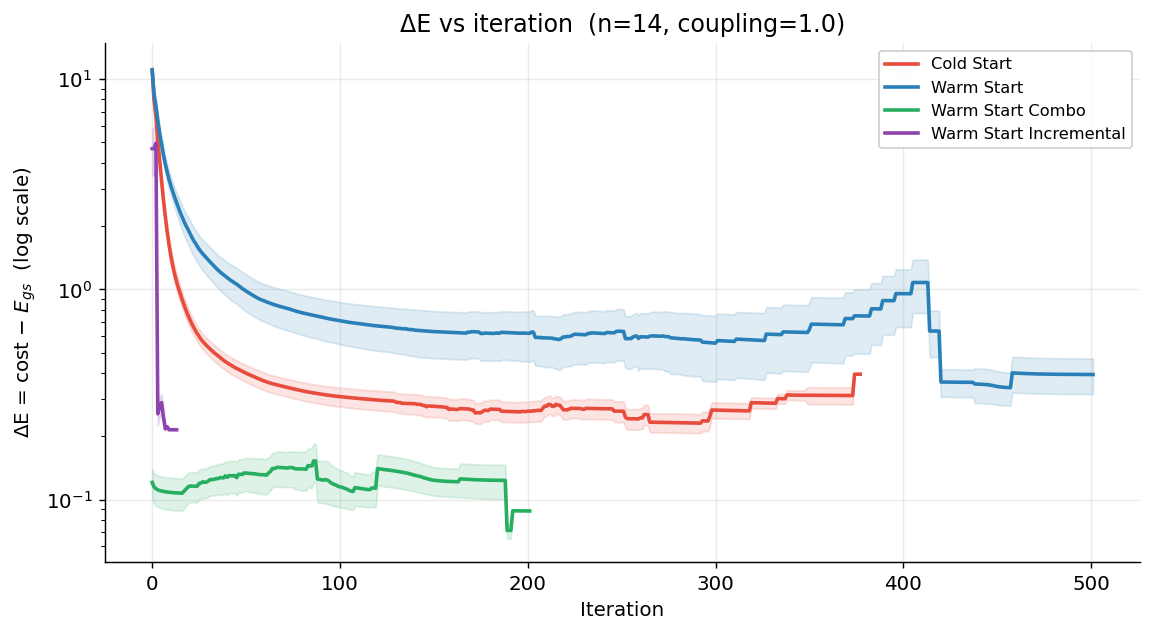}
    }\\[1.2em]

    \subfloat[$\beta$ parameter vs.\ system size $n$, $J = J_{\max}$.
    \label{fig:10chains_b}]{
        \includegraphics[width=0.95\columnwidth]{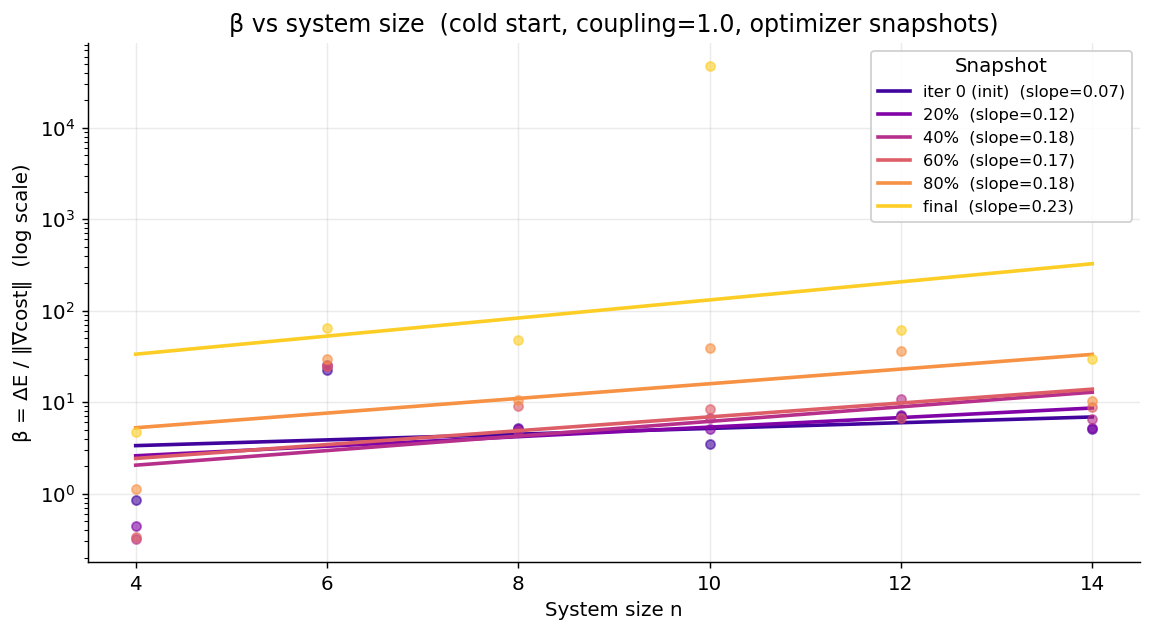}
    }\\[1.2em]

    \subfloat[Final fidelities  vs.\ system size $n$, $J = J_{\max}$.
    \label{fig:10chains_c}]{
        \includegraphics[width=0.95\columnwidth]{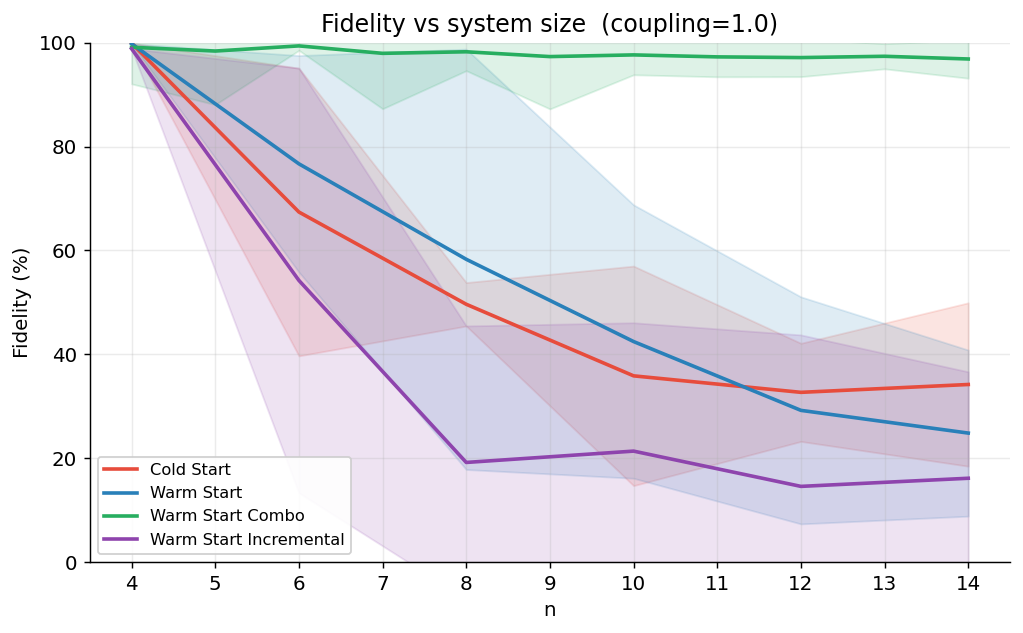}
    }

    \caption{The data gathered for the for a 14 qubit system, arranged in chain, at the coupling stage, $J_{max}$ for the different methods, cold start, warm start, warm start with incremental coupling increase, warm start combination (a) The energy difference between the observed and known ground state energy as the number of optimisation steps (b) The beta parameter $\beta = \frac{\Delta E}{\Vert\nabla C\Vert}$ which the difference in the cost function over the gradient of the cost function (c) The final fidelity of the different methods, Sample size = 100 }
    \label{fig:10chains}
\end{figure}

\begin{figure}[htbp!]
    \centering
    \includegraphics[width=0.8\linewidth]{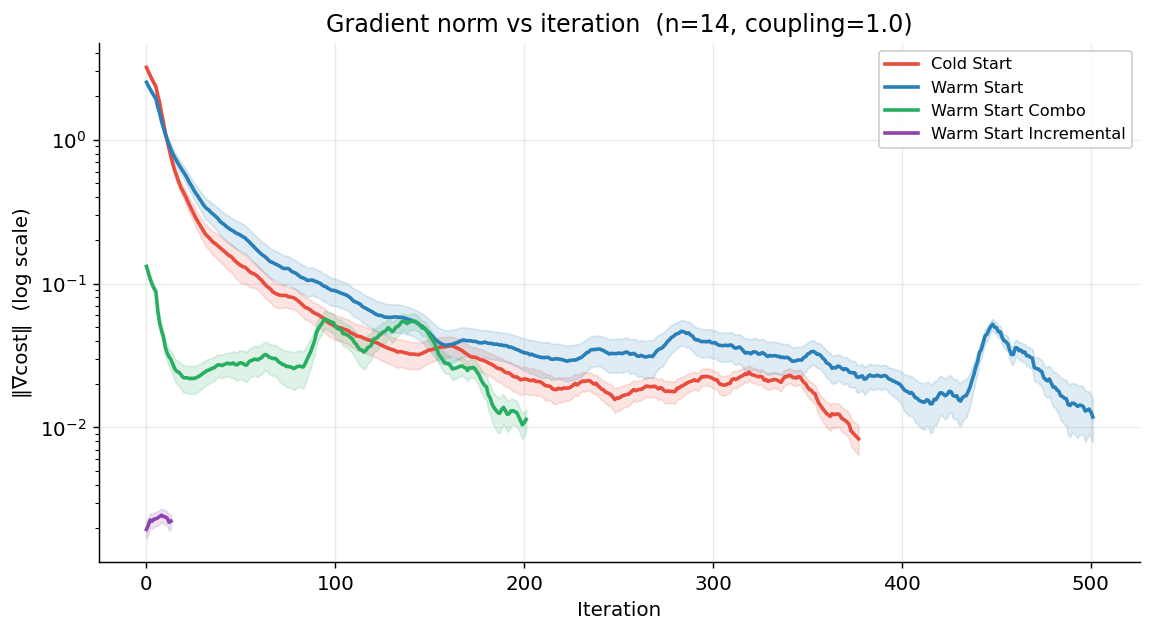}
    \caption{Gradient norm $||\nabla C||$ for Cold, Warm start by-size, Warm start Incremental, and Warm start Combination initialisation methods.}
    \label{fig:gradnorm}
\end{figure}

Compared to the cold start method, as well as the other two methods (warm start by size expansion and incremental ramp), the combination ramp method, which we will simply call the combo method, is able to achieve a high-fidelity many-body ground state. In each of the optimisation steps or iteration, the state evolves under the analog circuit, then we compute the expectation and gradient for the parameters. Following this, the curvature aware L-BFGS optimiser updates the parameters. The metrics energy, gradient, fidelity and $\beta$ are logged. The metrics for the $n=6$ complete graph and $n=14$ chain graph are shown in Fig.\ref{fig:6complete} and Fig.\ref{fig:10chains}   respectively. In the figures  \ref{fig:10chains} and \ref{fig:6complete}, the iteration count plotted is the number steps needed for the $n=14$ and $n=6$ qubit systems at maximum coupling strengths $J=J_{\text{max}}$ to reach the convergence. The grad norm $||\nabla C||$ in Fig.\ref{fig:gradnorm}, shows the decay of the gradient for the other methods is drastic, and combined with the energy difference, it can be seen that gradient drops, even before it reaches the ground state, thus suggesting it is stuck within some local area, and the fidelity drops.

\begin{figure}[!htbp]
    \centering

    \begin{subfigure}{0.5\textwidth}
        \centering
        \includegraphics[width=\textwidth]{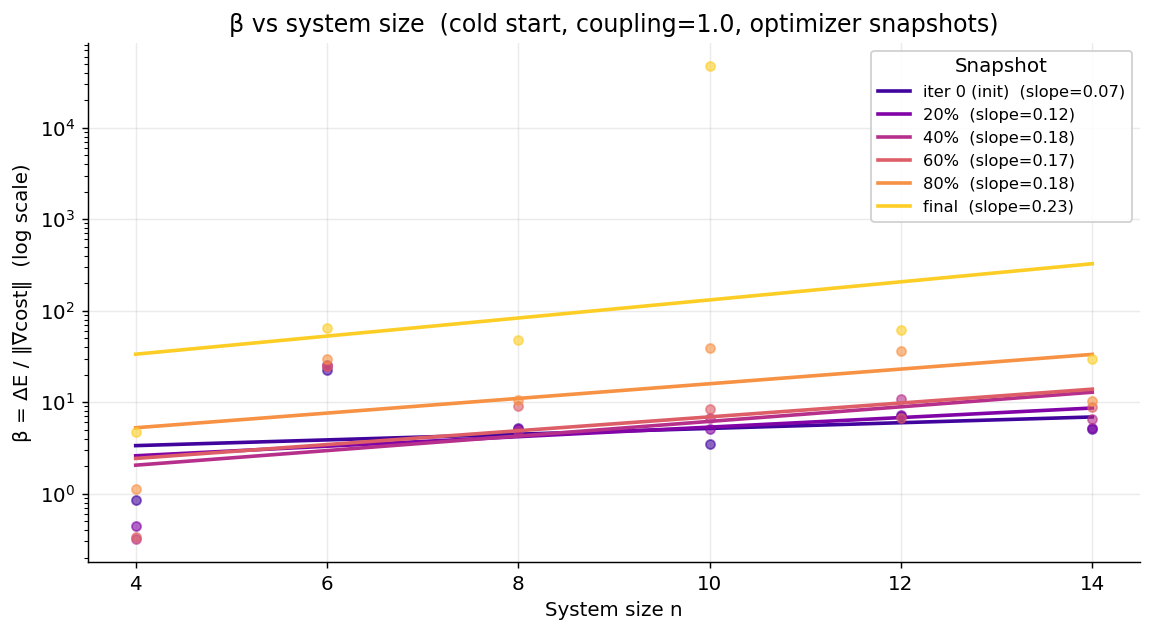}
        \caption{Beta parameter $\beta$ across various iterations for a cold start.}
        \label{fig:m1}
    \end{subfigure}
    \begin{subfigure}{0.5\textwidth}
        \centering
        \includegraphics[width=\textwidth]{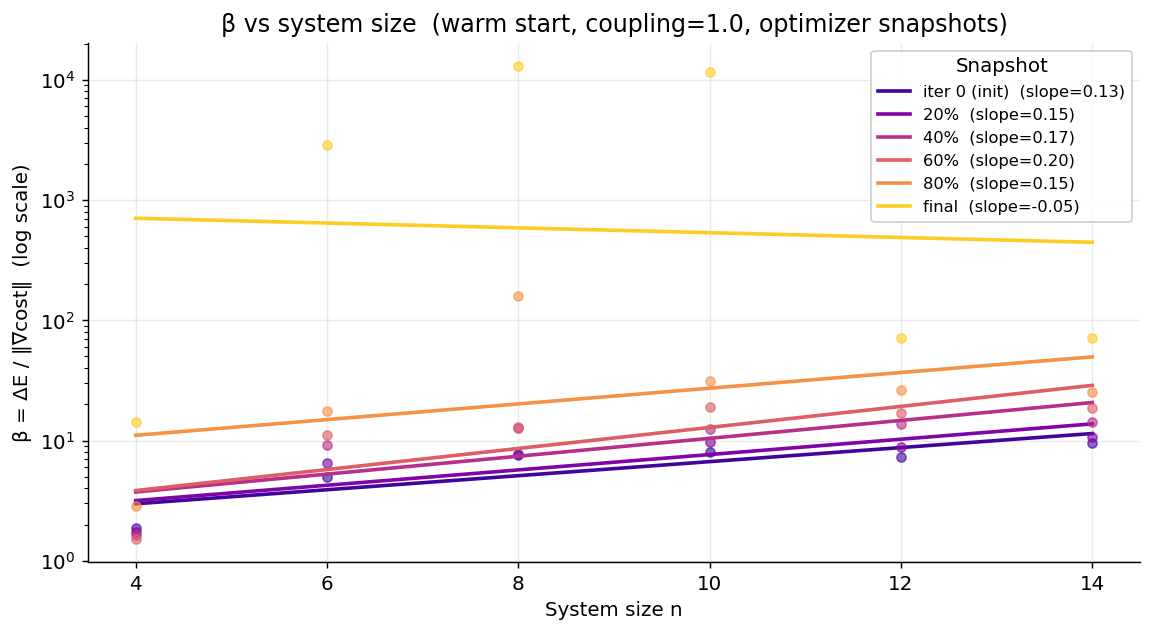}
        \caption{Beta parameter $\beta$ across various iterations for the Warm start by size expansion.}
        \label{fig:m2}
    \end{subfigure}
    \begin{subfigure}{0.5\textwidth}
        \centering
        \includegraphics[width=\textwidth]{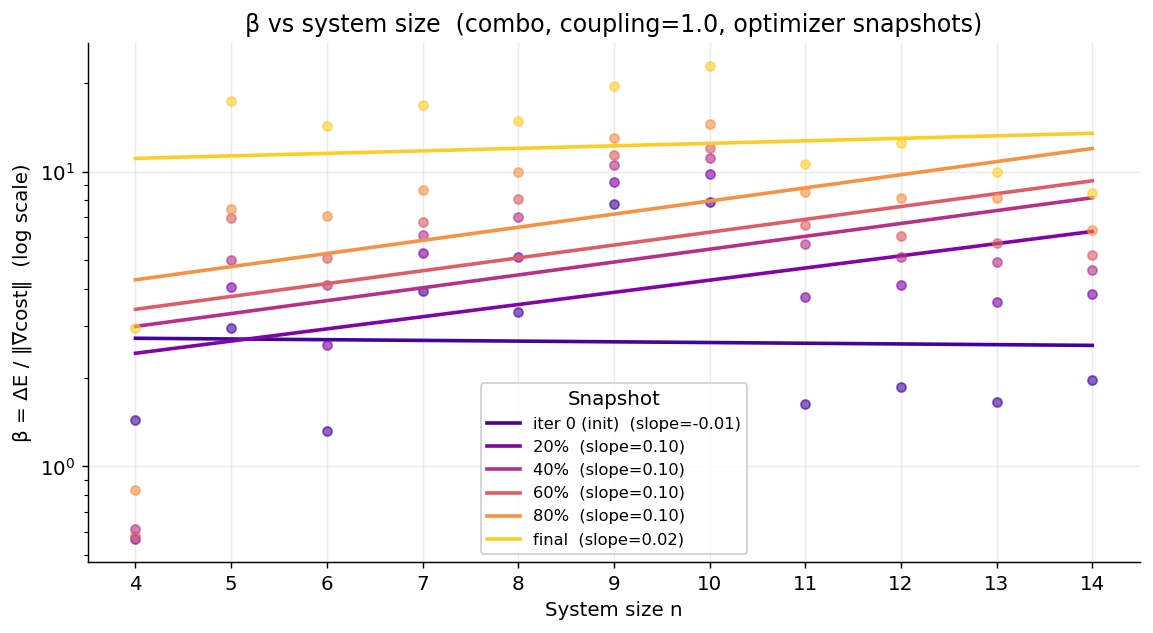}
        \caption{Beta parameter $\beta$ across various iterations for the Warm start combination.}
        \label{fig:m2}
    \end{subfigure}

    \caption{Comparison of the cold start and warm starts by size expansion and combination strategies.}
    \label{fig:betaperiteration}
\end{figure}

Comparison of the warm start by size and the combination methods in Fig.\ref{fig:betaperiteration} across iterations shows that the $\beta$ parameter decreases with increasing iteration number for the combination method, whereas for the warm start by size strategy, it grows exponentially.

It can also be noted that just the warm start or the incremental method by themselves do not show such notable behaviour, often performing the same as the cold or worse. Here we can only speculate the reasons for such a result. Often adding a single spin, for example, can drastically change the ground state of a quantum system. In Heisenberg spin models, the ground state after addition of a spin (going from even to odd, for example) can result in a very different entanglement structure, eg, the extra spin wants to bond (be, for example, entangled) with all other spins with which it is strongly connected. Thus the warm start only state may actually push the ground state to be reached quite  far. On the other hand, the incremental only method may be too slow, although it has the positive feature that the changes in the system's state is gradual enough as iteration progresses so that one can build on the previous steps of iteration. Thus only when warm start and incremental methods are combined, can we harness the power of the two techniques and achieve a high fidelity preparation of the ground state. 

\subsection{Abitrary Graphs}

\begin{figure}[!h]
    \centering

    \subfloat[Energy difference vs.\ iterations for $n=6$ complete graph, $J = J_{\max}$.%
    \label{fig:6CompleteE}]{
        \includegraphics[width=0.95\columnwidth]{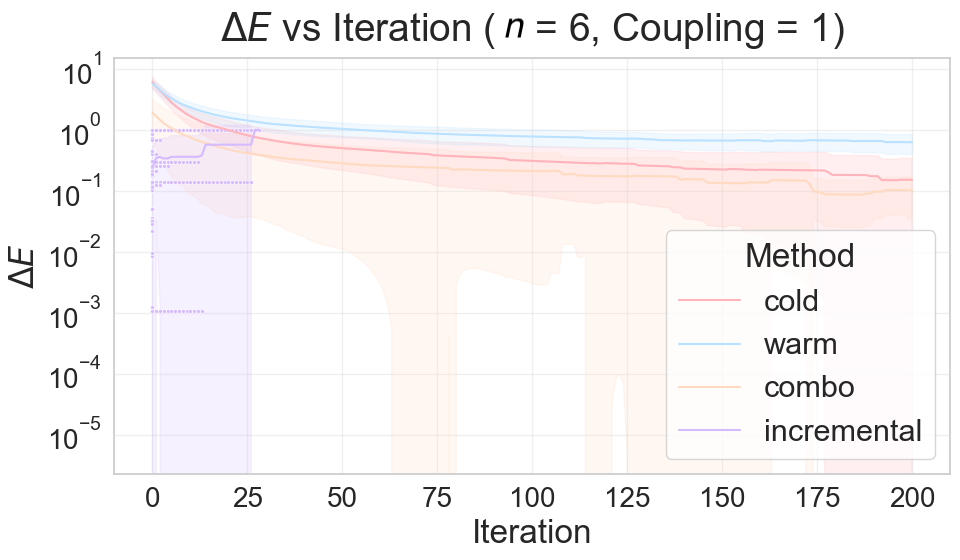}
    }\\[1.2em]

    \subfloat[$\beta$ parameter vs.\ iterations for $n=6$ complete graph, $J = J_{\max}$.%
    \label{fig:6Completeb}]{
        \includegraphics[width=0.95\columnwidth]{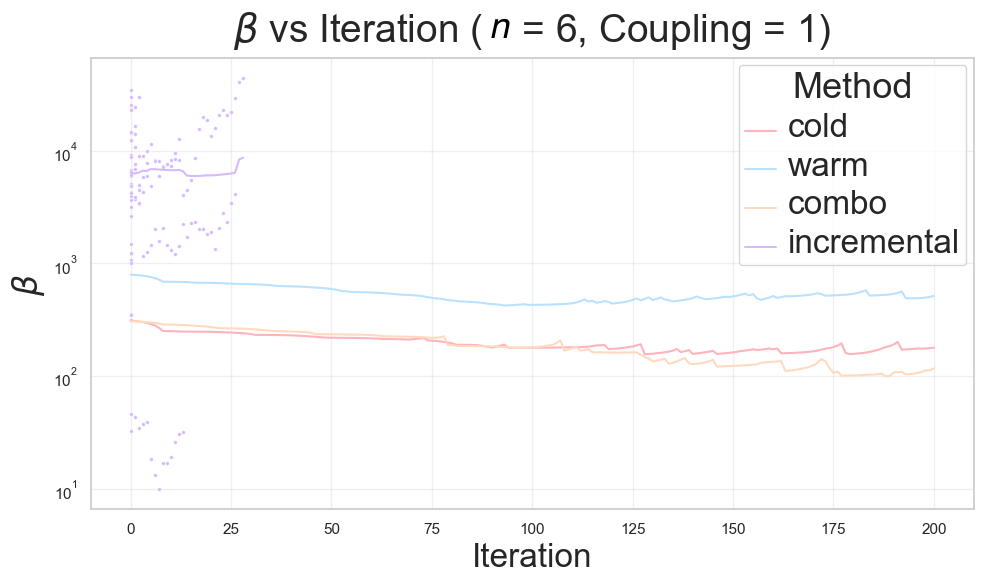}
    }\\[1.2em]

    \subfloat[Fidelity vs.\ iteration number for $n=6$ complete graph, $J = J_{\max}$.%
    \label{fig:6CompleteF}]{
        \includegraphics[width=0.95\columnwidth]{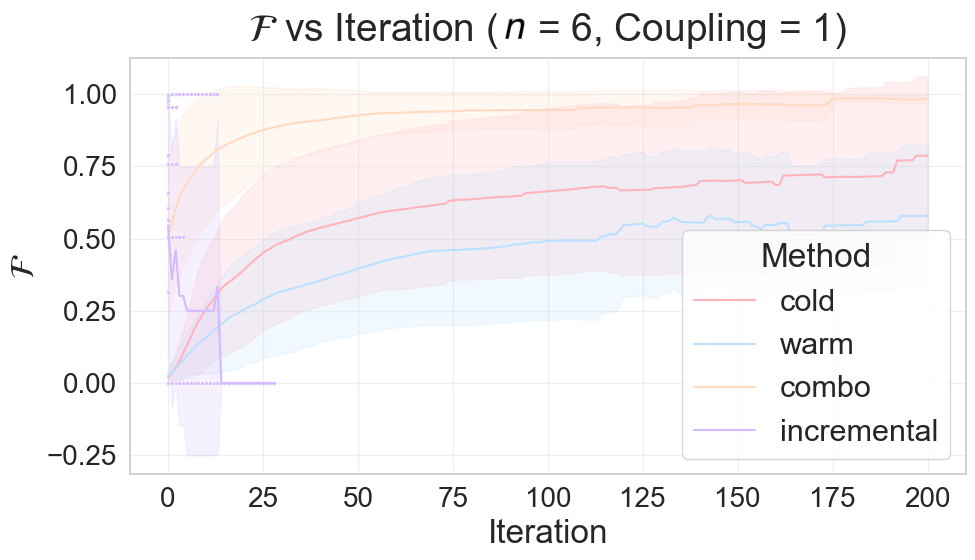}
    }

\caption{The data gathered for the for a 6 qubit system, complete graph at the coupling stage, $J_{max}$ for the different methods, cold start, warm start, warm start with incremental coupling increase, warm start combination (a) The energy difference between the observed and known ground state energy as the number of optimisation steps (b) The beta parameter $\beta = \frac{\Delta E}{\Vert\nabla C\Vert}$ which the difference in the cost function over the gradient of the cost function (c) The final fidelity of the different methods, Sample size = 100}
\label{fig:6complete}
\end{figure}

From Fig.\ref{fig:6complete} (ground states of complete graphs), we can clearly see that the energy convergence for the combo is much greater than the other techniques, the energy is always lower than $10^{-1}$, the $\beta$ parameter is also lower than the cold, warm and incremental, and of a noticeably higher fidelity $\sim 0.99$. The same pattern is also observed with the chains in Fig.\ref{fig:10chains}, and there is a more prominent difference between the combo and other methods, with the combo showing a much higher accuracy ($10^{-3}$) in the energy in comparison to the other methods. Fast stabilisation of $\beta$ parameter to a low value is seen (in fact, the gradient also does not decay for the combo as rapidly as the other methods). However, the most striking feature is the achieved fidelity -- it is $\sim 0.999$ for the combo method. 


\section{Conclusion}
In this paper, we have investigated a fast method to synthesise the ground state of given quantum many-body spin systems (specified by a problem Hamiltonian $H_p$) through the nonequilibrium dynamics in a quantum simulator. The initial state of the simulator is taken to be some fiducial state (eg $|+\rangle^{\otimes n}$ or $|0\rangle^{\otimes n}$) and the dynamics is given by a solver Hamiltonian $H_s$ assumed realisable in the simulator. One can also think that $H_p$ poses a quantum problem which is solved by the solver Hamiltonian $H_s$. The solver Hamiltonian only acts for one unit of time, when the couplings are kept to be of the order of unity. This preparation method is then much faster than both the natural relaxation or adiabatic dynamics of the many-body system, which, by definition cannot be faster than the inverse gap $1/\Delta$.   Note that $H_s\neq H_p$, which is the very fact that allows one to reach the ground state (an eigenstate of $H_p$) with $H_s$.

To obtain the optimal $H_s$ for a given $H_p$ we have used the energy as a cost function for minimization, with techniques similar to variational quantum algorithms, but simulated everything classically for up to $n=6$ for graphs and $n=14$ for chains. In this context, we have also found that a technique the combines warm starts with slow increments of coupling to added qubits, the combo method, is required to get a good fidelity of preparation of the ground state. 
Presumably, with more powerful computers and a greater investment of time, the optimal $H_s$ can be found for a few more qubits. However, we need to invest the computing power and time to find the optimal $H_s$ for a given $H_p$ {\em only once}. Then it can be catalogued, and the ground state of any specific $H_p$ can henceforth be prepared rapidly in the quantum simulator with the corresponding optimal $H_s$ (note that ``knowing'' $\ket{GS(H_p(\bf{J}_p))}$ by exact diagonalization does not tell us how to prepare it fast in a simulator). The ground state of several (i.e, $n=6-14$), but still limited, number of spins in itself is useful as a starting resource of several investigations in a quantum simulator which extrapolate to larger sizes eg, just a bond quench to connect two ground states, or multiple bond quenches to connect several ground states or add copies to generate higher dimensional states, more complex graphs, and measure the resulting (classically non-simulable) dynamics on a quantum simulator.

However, beyond the above use, purely as a classical procedure for finding the method to prepare a quantum ground state in an analog simulator where $H_s$ can be realized, for larger lattice sizes, we can also think of our procedure of finding $H_s$ as a hybrid quantum-classical algorithm for an analog quantum processor. Some differences with respect to the types of variational quantum eigensolvers commonly studied are that neither is this a parametrized quantum circuit comprised of a series of gates \cite{bharti2022noisy,tilly2022variational}, nor is it comprised of  several smaller Hamiltonian components of $H_p$ itself \cite{wecker2015progress,hadfield2019quantum,kokail2019self,lyu2020accelerated,feulner2022variational,grimsley2023adaptive}. With respect to methodologies such as Quantum Approximate Optimization Algorithm or its generalized version Quantum Alternating Operator Ansatz \cite{hadfield2019quantum}, we differ by not alternating $H_p$ itself with a mixer. Additionally, it is a very shallow circuit with a single layer. It is close in spirit to designing multi-qubit unitary operations with time constant Hamiltonians \cite{PhysRevA.111.052618}, except that here we are aiming for a specific state (the ground state) to be obtained by a similar evolution. However, when our methodology is used as a hybrid quantum-classical algorithm, it incorporates a very positive aspect of usual variational quantum algorithms: the optimal
$H_s$ is searched for via a variational approach, where the couplings of $H_s$ are optimised to minimise the cost function $\bra{GS(H_p(\bf{J}_p))}H_p\ket{GS(H_p(\bf{J}_p))}$, which is simply the energy of the dynamically produced state  with respect to the problem Hamiltonian. The learning of the parameters is thus accomplished by an efficiently measurable cost function, as the Hamiltonian $H_p$ consists as best of two-body terms (measurable as two body correlation functions), of which there are only about $O(n^2)$ terms, although for energy accuracy of $\epsilon$, about $\sim 1/\epsilon^2$ measurement shots have to be included in the quantum version for determining each correlation function. The proposed methodology of optimization in this paper: the warm start and incremental method combination ramp shows the best performance while the $\beta$ parameter suggests  that the gradient does not decay fast, hence energies closer to the ground state can be reached. It will be of interest to apply our methodology of evolution via a variationally optimized time constant, but problem inspired (but obviously non-commuting and different) solver Hamiltonian $H_s$ to other systems to obtain thrie ground states in a short time.  

\section*{acknowledgments}
PT acknowledges support from the EPSRC Centre for Doctoral Training in Delivering Quantum Technologies
Grant EP/S021582/1. SB acknowledges UK Research
and Innovation (UKRI) Grant No. EP/R029075/1 and EPSRC-SFI project EP/X039889/1(GeQuantumBus). We thank Abolfazl Bayat for useful discussions. 


\bibliography{ref}

@article{preskill2018quantum,
  title={Quantum computing in the NISQ era and beyond},
  author={Preskill, John},
  journal={Quantum},
  volume={2},
  pages={79},
  year={2018},
  doi={10.22331/q-2018-08-06-79}
}

@book{sachdev2011quantum,
  title={Quantum Phase Transitions},
  author={Sachdev, Subir},
  year={2011},
  publisher={Cambridge University Press},
  edition={2nd}
}

@article{vojta2003quantum,
  title={Quantum phase transitions},
  author={Vojta, Matthias},
  journal={Reports on Progress in Physics},
  volume={66},
  number={12},
  pages={2069},
  year={2003},
  doi={10.1088/0034-4885/66/12/R01}
}

@article{greiner2002quantum,
  title={Quantum phase transition from a superfluid to a Mott insulator in a gas of ultracold atoms},
  author={Greiner, Markus and Mandel, Olaf and Esslinger, Tilman and Hänsch, Theodor W and Bloch, Immanuel},
  journal={Nature},
  volume={415},
  number={6867},
  pages={39--44},
  year={2002},
  doi={10.1038/415039a}
}

@article{li2025low,
  title={Low lying excited states quantum entanglement and continuous quantum phase transitions},
  author={Li, Yan-Chao and Zhou, Yuan-Hang and Zhang, Yuan and Lin, Hai-Qing},
  journal={Scientific Reports},
  volume={15},
  pages={6277},
  year={2025},
  doi={10.1038/s41598-025-90248-0},
  note={Article number: 6277}
}

@article{ward2009preparation,
  title={Preparation of many-body states for quantum simulation},
  author={Ward, Nicholas J and Kassal, Ivan and Aspuru-Guzik, Al{\'a}n},
  journal={The Journal of chemical physics},
  volume={130},
  number={19},
  year={2009},
  publisher={AIP Publishing}
}

@article{Calzetta_2018,
   title={Not-quite-free shortcuts to adiabaticity},
   volume={98},
   ISSN={2469-9934},
   url={http://dx.doi.org/10.1103/PhysRevA.98.032107},
   DOI={10.1103/physreva.98.032107},
   number={3},
   journal={Physical Review A},
   publisher={American Physical Society (APS)},
   author={Calzetta, Esteban},
   year={2018},
   month=sep }

@article{B_rligea_2025,
   title={Scalability challenges in variational quantum optimization under stochastic noise},
   volume={112},
   ISSN={2469-9934},
   DOI={10.1103/rgyh-8xw8},
   number={3},
   journal={Physical Review A},
   publisher={American Physical Society (APS)},
   author={Bärligea, Adelina and Poggel, Benedikt and Lorenz, Jeanette Miriam},
   year={2025},
   month=sep }

@misc{korpas2025undecidableproblemsassociatedvariational,
      title={Undecidable problems associated with variational quantum algorithms}, 
      author={Georgios Korpas and Vyacheslav Kungurtsev and Jakub Mareček},
      year={2025},
      eprint={2503.23723},
      archivePrefix={arXiv},
      primaryClass={quant-ph},
      url={https://arxiv.org/abs/2503.23723}, 
}

@misc{teoh2019machinelearningdesigntrappedion,
      title={Machine learning design of a trapped-ion quantum spin simulator}, 
      author={Yi Hong Teoh and Marina Drygala and Roger G. Melko and Rajibul Islam},
      year={2019},
      eprint={1910.02496},
      archivePrefix={arXiv},
      primaryClass={quant-ph},
      url={https://arxiv.org/abs/1910.02496}, 
}

@article{feynman1982simulating,
  title={Simulating Physics with Computers},
  author={Feynman, Richard P},
  journal={International Journal of Theoretical Physics},
  volume={21},
  number={6/7},
  year={1982}
}

@article{macdonell2021analog,
  title={Analog quantum simulation of chemical dynamics},
  author={MacDonell, Ryan J and Dickerson, Claire E and Birch, Clare JT and Kumar, Alok and Edmunds, Claire L and Biercuk, Michael J and Hempel, Cornelius and Kassal, Ivan},
  journal={Chemical science},
  volume={12},
  number={28},
  pages={9794--9805},
  year={2021},
  publisher={Royal Society of Chemistry}
}

@article{arguello2019analogue,
  title={Analogue quantum chemistry simulation},
  author={Arg{\"u}ello-Luengo, Javier and Gonz{\'a}lez-Tudela, Alejandro and Shi, Tao and Zoller, Peter and Cirac, J Ignacio},
  journal={Nature},
  volume={574},
  number={7777},
  pages={215--218},
  year={2019},
  publisher={Nature Publishing Group UK London}
}

@article{gibbs2025exploiting,
  title={Exploiting symmetries in nuclear Hamiltonians for ground state preparation},
  author={Gibbs, Joe and Holmes, Zo{\"e} and Stevenson, Paul},
  journal={Quantum Machine Intelligence},
  volume={7},
  number={1},
  pages={14},
  year={2025},
  publisher={Springer}
}

@article{kokail2019self,
  title={Self-verifying variational quantum simulation of lattice models},
  author={Kokail, Christian and Maier, Christine and van Bijnen, Rick and Brydges, Tiff and Joshi, Manoj K and Jurcevic, Petar and Muschik, Christine A and Silvi, Pietro and Blatt, Rainer and Roos, Christian F and others},
  journal={Nature},
  volume={569},
  number={7756},
  pages={355--360},
  year={2019},
  publisher={Nature Publishing Group UK London}
}

@article{feulner2022variational,
  title={Variational quantum eigensolver ansatz for the j 1-j 2-model},
  author={Feulner, Verena and Hartmann, Michael J},
  journal={Physical Review B},
  volume={106},
  number={14},
  pages={144426},
  year={2022},
  publisher={APS}
}

@article{larocca2025barren,
  title={Barren plateaus in variational quantum computing},
  author={Larocca, Martin and Thanasilp, Supanut and Wang, Samson and Sharma, Kunal and Biamonte, Jacob and Coles, Patrick J and Cincio, Lukasz and McClean, Jarrod R and Holmes, Zo{\"e} and Cerezo, Marco},
  journal={Nature Reviews Physics},
  volume={7},
  pages={174-189},
  year={2025},
  publisher={Nature Publishing Group UK London}
}

@article{park2024hamiltonian,
  title={Hamiltonian variational ansatz without barren plateaus},
  author={Park, Chae-Yeun and Killoran, Nathan},
  journal={Quantum},
  volume={8},
  pages={1239},
  year={2024},
  publisher={Verein zur F{\"o}rderung des Open Access Publizierens in den Quantenwissenschaften}
}

@article{wecker2015progress,
  title={Progress towards practical quantum variational algorithms},
  author={Wecker, Dave and Hastings, Matthew B and Troyer, Matthias},
  journal={Physical Review A},
  volume={92},
  number={4},
  pages={042303},
  year={2015},
  publisher={APS}
}

@article{puig2025variational,
  title={Variational quantum simulation: a case study for understanding warm starts},
  author={Puig, Ricard and Drudis, Marc and Thanasilp, Supanut and Holmes, Zo{\"e}},
  journal={PRX Quantum},
  volume={6},
  number={1},
  pages={010317},
  year={2025},
  publisher={APS}
}

@article{hadfield2019quantum,
  title={From the quantum approximate optimization algorithm to a quantum alternating operator ansatz},
  author={Hadfield, Stuart and Wang, Zhihui and O’gorman, Bryan and Rieffel, Eleanor G and Venturelli, Davide and Biswas, Rupak},
  journal={Algorithms},
  volume={12},
  number={2},
  pages={34},
  year={2019},
  publisher={MDPI}
}

@article{grimsley2023adaptive,
  title={Adaptive, problem-tailored variational quantum eigensolver mitigates rough parameter landscapes and barren plateaus},
  author={Grimsley, Harper R and Barron, George S and Barnes, Edwin and Economou, Sophia E and Mayhall, Nicholas J},
  journal={npj Quantum Information},
  volume={9},
  number={1},
  pages={19},
  year={2023},
  publisher={Nature Publishing Group UK London}
}

@article{dborin2022matrix,
  title={Matrix product state pre-training for quantum machine learning},
  author={Dborin, James and Barratt, Fergus and Wimalaweera, Vinul and Wright, Lewis and Green, Andrew G},
  journal={Quantum Science and Technology},
  volume={7},
  number={3},
  pages={035014},
  year={2022},
  publisher={IOP Publishing}
}

@article{rudolph2023synergistic,
  title={Synergistic pretraining of parametrized quantum circuits via tensor networks},
  author={Rudolph, Manuel S and Miller, Jacob and Motlagh, Danial and Chen, Jing and Acharya, Atithi and Perdomo-Ortiz, Alejandro},
  journal={Nature Communications},
  volume={14},
  number={1},
  pages={8367},
  year={2023},
  publisher={Nature Publishing Group UK London}
}

@article{mele2022avoiding,
  title={Avoiding barren plateaus via transferability of smooth solutions in a Hamiltonian variational ansatz},
  author={Mele, Antonio A and Mbeng, Glen B and Santoro, Giuseppe E and Collura, Mario and Torta, Pietro},
  journal={Physical Review A},
  volume={106},
  number={6},
  pages={L060401},
  year={2022},
  publisher={APS}
}

@article{lyu2020accelerated,
  title={Accelerated variational algorithms for digital quantum simulation of many-body ground states},
  author={Lyu, Chufan and Montenegro, Victor and Bayat, Abolfazl},
  journal={Quantum},
  volume={4},
  pages={324},
  year={2020},
  publisher={Verein zur F{\"o}rderung des Open Access Publizierens in den Quantenwissenschaften}
}

@article{lyu2023symmetry,
  title={Symmetry enhanced variational quantum spin eigensolver},
  author={Lyu, Chufan and Xu, Xusheng and Yung, Man-Hong and Bayat, Abolfazl},
  journal={Quantum},
  volume={7},
  pages={899},
  year={2023},
  publisher={Verein zur F{\"o}rderung des Open Access Publizierens in den Quantenwissenschaften}
}

@article{jaderberg2025variational,
  title={Variational preparation of normal matrix product states on quantum computers},
  author={Jaderberg, Ben and Pennington, George and Marshall, Kate V and Anderson, Lewis W and Agarwal, Abhishek and Lindoy, Lachlan P and Rungger, Ivan and Mensa, Stefano and Crain, Jason},
  journal={arXiv preprint arXiv:2503.09683},
  year={2025}
}

@article{PhysRevA.111.052618,
  title = {Geodesic algorithm for unitary gate design with time-independent Hamiltonians},
  author = {Lewis, Dylan and Wiersema, Roeland and Carrasquilla, Juan and Bose, Sougato},
  journal = {Phys. Rev. A},
  volume = {111},
  issue = {5},
  pages = {052618},
  numpages = {15},
  year = {2025},
  month = {May},
  publisher = {American Physical Society},
  doi = {10.1103/PhysRevA.111.052618},
  url = {https://link.aps.org/doi/10.1103/PhysRevA.111.052618}
}

@article{bespalova2021quantum,
  title={Quantum simulation and ground state preparation for the honeycomb Kitaev model},
  author={Bespalova, Tatiana A and Kyriienko, Oleksandr},
  journal={arXiv preprint arXiv:2109.13883},
  year={2021}
}

@article{barmettler2010quantum,
  title={Quantum quenches in the anisotropic spin-Heisenberg chain: different approaches to many-body dynamics far from equilibrium},
  author={Barmettler, Peter and Punk, Matthias and Gritsev, Vladimir and Demler, Eugene and Altman, Ehud},
  journal={New Journal of Physics},
  volume={12},
  number={5},
  pages={055017},
  year={2010},
  publisher={IOP Publishing}
}

@article{sels2022bath,
  title={Bath-induced delocalization in interacting disordered spin chains},
  author={Sels, Dries},
  journal={Physical Review B},
  volume={106},
  number={2},
  pages={L020202},
  year={2022},
  publisher={APS}
}

@article{bayat2010entanglement,
  title={Entanglement routers using macroscopic singlets},
  author={Bayat, Abolfazl and Bose, Sougato and Sodano, Pasquale},
  journal={Physical review letters},
  volume={105},
  number={18},
  pages={187204},
  year={2010},
  publisher={APS}
}

@article{umer2025probing,
  title={Probing the limits of variational quantum algorithms for nonlinear ground states on real quantum hardware: The effects of noise},
  author={Umer, Muhammad and Mastorakis, Eleftherios and Evangelou, Sofia and Angelakis, Dimitris G},
  journal={Physical Review A},
  volume={111},
  number={1},
  pages={012626},
  year={2025},
  publisher={APS}
}

@article{lubasch2011adiabatic,
   title={Adiabatic Preparation of a Heisenberg Antiferromagnet Using an Optical Superlattice},
   volume={107},
   ISSN={1079-7114},
   url={http://dx.doi.org/10.1103/PhysRevLett.107.165301},
   DOI={10.1103/physrevlett.107.165301},
   number={16},
   journal={Physical Review Letters},
   publisher={American Physical Society (APS)},
   author={Lubasch, Michael and Murg, Valentin and Schneider, Ulrich and Cirac, J. Ignacio and Bañuls, Mari-Carmen},
   year={2011},
   month=Oct }

@article{farooq2015adiabatic,
   title={Adiabatic many-body state preparation and information transfer in quantum dot arrays},
   volume={91},
   ISSN={1550-235X},
   url={http://dx.doi.org/10.1103/PhysRevB.91.134303},
   DOI={10.1103/physrevb.91.134303},
   number={13},
   journal={Physical Review B},
   publisher={American Physical Society (APS)},
   author={Farooq, Umer and Bayat, Abolfazl and Mancini, Stefano and Bose, Sougato},
   year={2015},
   month=Apr }

@article{bharti2022noisy,
  title={Noisy intermediate-scale quantum algorithms},
  author={Bharti, Kishor and Cervera-Lierta, Alba and Kyaw, Thi Ha and Haug, Tobias and Alperin-Lea, Sumner and Anand, Abhinav and Degroote, Matthias and Heimonen, Hermanni and Kottmann, Jakob S and Menke, Tim and others},
  journal={Reviews of Modern Physics},
  volume={94},
  number={1},
  pages={015004},
  year={2022},
  publisher={APS}
}

@article{tilly2022variational,
  title={The variational quantum eigensolver: a review of methods and best practices},
  author={Tilly, Jules and Chen, Hongxiang and Cao, Shuxiang and Picozzi, Dario and Setia, Kanav and Li, Ying and Grant, Edward and Wossnig, Leonard and Rungger, Ivan and Booth, George H and others},
  journal={Physics Reports},
  volume={986},
  pages={1--128},
  year={2022},
  publisher={Elsevier}
}

@misc{ryabinkin2018qubit,
      title={Iterative Qubit Coupled Cluster approach with efficient screening of generators}, 
      author={Ilya G. Ryabinkin and Robert A. Lang and Scott N. Genin and Artur F. Izmaylov},
      year={2019},
      eprint={1906.11192},
      archivePrefix={arXiv},
      primaryClass={quant-ph},
      url={https://arxiv.org/abs/1906.11192}, 
}

@article{kunitski2019double,
  title={Double-slit photoelectron interference in strong-field ionization of the neon dimer},
  author={Kunitski, Maksim and Eicke, Nicolas and Huber, Pia and K{\"o}hler, Jonas and Zeller, Stefan and Voigtsberger, J{\"o}rg and Schlott, Nikolai and Henrichs, Kevin and Sann, Hendrik and Trinter, Florian and Schmidt, Lothar Ph. H. and Kalinin, Anton and Sch{\"o}ffler, Markus S. and Jahnke, Till and Lein, Manfred and D{\"o}rner, Reinhard},
  journal={Nature Communications},
  volume={10},
  number={1},
  pages={1--8},
  year={2019},
  doi={10.1038/s41467-018-07882-8},
  publisher={Nature Publishing Group}
}

@article{PhysRevResearch.2.023074,
  title = {Quantum optimization with a novel Gibbs objective function and ansatz architecture search},
  author = {Li, Li and Fan, Minjie and Coram, Marc and Riley, Patrick and Leichenauer, Stefan},
  journal = {Phys. Rev. Res.},
  volume = {2},
  issue = {2},
  pages = {023074},
  numpages = {10},
  year = {2020},
  month = {Apr},
  publisher = {American Physical Society},
  doi = {10.1103/PhysRevResearch.2.023074},
  url = {https://link.aps.org/doi/10.1103/PhysRevResearch.2.023074}
}

@article{du2022quantum,
  title={Quantum circuit architecture search for variational quantum algorithms},
  author={Du, Yuxuan and Huang, Tao and You, Shan and Hsieh, Min-Hsiu and Tao, Dacheng},
  journal={npj Quantum Information},
  volume={8},
  pages={62},
  year={2022},
  doi={10.1038/s41534-022-00570-y},
  publisher={Nature Publishing Group}
}

@article{PRXQuantum.2.010324,
  title = {Machine Learning of Noise-Resilient Quantum Circuits},
  author = {Cincio, Lukasz and Rudinger, Kenneth and Sarovar, Mohan and Coles, Patrick J.},
  journal = {PRX Quantum},
  volume = {2},
  issue = {1},
  pages = {010324},
  numpages = {19},
  year = {2021},
  month = {Feb},
  publisher = {American Physical Society},
  doi = {10.1103/PRXQuantum.2.010324},
  url = {https://link.aps.org/doi/10.1103/PRXQuantum.2.010324}
}

@article{kuo2021quantum,
  title={Quantum architecture search via deep reinforcement learning},
  author={Kuo, En-Jui and Fang, Yao-Lung L. and Chen, Samuel Yen-Chi},
  journal={arXiv preprint arXiv:2104.07715},
  year={2021},
  url={https://arxiv.org/abs/2104.07715},
  note={Submitted April 15, 2021}
}

@article{altares2021automatic,
  title={Automatic design of quantum feature maps},
  author={Altares-L{\'o}pez, Sergio and Ribeiro, Angela and Garc{\'\i}a-Ripoll, Juan Jos{\'e}},
  journal={Quantum Science and Technology},
  volume={6},
  number={4},
  pages={045015},
  year={2021},
  doi={10.1088/2058-9565/ac1ab1},
  publisher={IOP Publishing}
}

@article{huang2022robust,
  title={Robust resource-efficient quantum variational ansatz through an evolutionary algorithm},
  author={Huang, Yuhan and Li, Qingyu and Hou, Xiaokai and Wu, Rebing and Yung, Man‑Hong and Bayat, Abolfazl and Wang, Xiaoting},
  journal={Physical Review A},
  volume={105},
  number={5},
  pages={052414},
  year={2022},
  doi={10.1103/PhysRevA.105.052414},
  publisher={American Physical Society}
}

@article{sun2024quantum,
   title={Quantum Architecture Search with Unsupervised Representation Learning},
   volume={10},
   ISSN={2521-327X},
   url={http://dx.doi.org/10.22331/q-2026-02-03-1994},
   DOI={10.22331/q-2026-02-03-1994},
   journal={Quantum},
   publisher={Verein zur Forderung des Open Access Publizierens in den Quantenwissenschaften},
   author={Sun, Yize and Wu, Zixin and Tresp, Volker and Ma, Yunpu},
   year={2026},
   month=Feb, pages={1994} }

@article{han2024multilevel,
  title={Multilevel variational spectroscopy using a programmable quantum simulator},
  author={Han, Zhikun and Lyu, Chufan and Zhou, Yuxuan and Yuan, Jiahao and Chu, Ji and Nuerbolati, Wuerkaixi and Jia, Hao and Nie, Lifu and Wei, Weiwei and Yang, Zusheng and Zhang, Libo and Zhang, Ziyan and Hu, Chang-Kang and Hu, Ling and Li, Jian and Tan, Dian and Bayat, Abolfazl and Liu, Song and Yan, Fei and Yu, Dapeng},
  journal={Physical Review Research},
  volume={6},
  pages={013015},
  year={2024},
  doi={10.1103/PhysRevResearch.6.013015},
  publisher={American Physical Society}
}

@article{cerezo2021variational,
  title = {Variational quantum algorithms},
  author = {Cerezo, Marco and Arrasmith, Andrew and Babbush, Ryan and Benjamin, Simon C. and Endo, Suguru and Fujii, Keisuke and McClean, Jarrod R. and Mitarai, Kosuke and Yuan, Xiao and Cincio, Lukasz and Coles, Patrick J.},
  journal = {Nature Reviews Physics},
  volume = {3},
  pages = {625--644},
  year = {2021},
  doi = {10.1038/s42254-021-00348-9},
  publisher = {Nature Publishing Group}
}

@article{peruzzo2014variational,
  title = {A variational eigenvalue solver on a photonic quantum processor},
  author = {Peruzzo, Alberto and McClean, Jarrod and Shadbolt, Peter and Yung, Man-Hong and Zhou, Xiao-Qi and Love, Peter J. and Aspuru-Guzik, Alan and O'Brien, Jeremy L.},
  journal = {Nature Communications},
  volume = {5},
  pages = {4213},
  year = {2014},
  doi = {10.1038/ncomms5213},
  publisher = {Nature Publishing Group}
}

@article{arute2020hartree,
  title = {Hartree-Fock on a superconducting qubit quantum computer},
  author = {Arute, Frank and Arya, Kunal and Babbush, Ryan and Bacon, Dave and Bardin, Joseph C. and Barends, Rami and Boixo, Sergio and Broughton, Michael and Buckley, Bob B. and Buell, David A. and {\em et al.}},
  journal = {Science},
  volume = {369},
  number = {6507},
  pages = {1084--1089},
  year = {2020},
  doi = {10.1126/science.abb9811},
  publisher = {American Association for the Advancement of Science}
}

@article{nature_computational_science_2022,
   title={Experimental quantum adversarial learning with programmable superconducting qubits},
   volume={2},
   ISSN={2662-8457},
   url={http://dx.doi.org/10.1038/s43588-022-00351-9},
   DOI={10.1038/s43588-022-00351-9},
   number={11},
   journal={Nature Computational Science},
   publisher={Springer Science and Business Media LLC},
   author={Ren, Wenhui and Li, Weikang and Xu, Shibo and Wang, Ke and Jiang, Wenjie and Jin, Feitong and Zhu, Xuhao and Chen, Jiachen and Song, Zixuan and Zhang, Pengfei and Dong, Hang and Zhang, Xu and Deng, Jinfeng and Gao, Yu and Zhang, Chuanyu and Wu, Yaozu and Zhang, Bing and Guo, Qiujiang and Li, Hekang and Wang, Zhen and Biamonte, Jacob and Song, Chao and Deng, Dong-Ling and Wang, H.},
   year={2022},
   month=Nov, pages={711–717} }

@article{schuld2021effect,
  title={Effect of data encoding on the expressive power of variational quantum‑machine‑learning models},
  author={Schuld, Maria and Sweke, Ryan and Meyer, Johannes Jakob},
  journal={Physical Review A},
  volume={103},
  number={3},
  pages={032430},
  year={2021},
  doi={10.1103/PhysRevA.103.032430},
  publisher={American Physical Society}
}

@article{wu2025enhancing,
  title={Enhancing the reachability of variational quantum algorithms via input‑state design},
  author={Wu, Shaojun and Jin, Shan and Bayat, Abolfazl and Wang, Xiaoting},
  journal={arXiv preprint arXiv:2510.26379},
  year={2025},
  url={https://arxiv.org/abs/2510.26379},
  note={Last revised 30 Oct 2025}
}

@article{cunningham2025investigating,
  title={Investigating and mitigating barren plateaus in variational quantum circuits: a survey},
  author={Cunningham, Jack and Zhuang, Jun},
  journal={Quantum Information Processing},
  volume={24},
  number={48},
  year={2025},
  publisher={Springer},
  doi={10.1007/s11128-025-04665-1}
}

@article{ragone2024lie,
  title={A Lie algebraic theory of barren plateaus for deep parameterized quantum circuits},
  author={Ragone, Michael and Bakalov, Bojko N and Sauvage, Fr{\'e}d{\'e}ric and Kemper, Alexander F and Ortiz Marrero, Carlos and Larocca, Martin and Cerezo, M},
  journal={Nature Communications},
  volume={15},
  number={1},
  pages={7172},
  year={2024},
  publisher={Nature Publishing Group},
  doi={10.1038/s41467-024-49909-3}
}

@article{chen2024taming,
  title={Taming Barren Plateaus in Arbitrary Parameterized Quantum Circuits Without Sacrificing Expressibility},
  author={Chen, Zhenyu and Shao, Yuguo and Liu, Zhengwei and Wei, Zhaohui},
  journal={arXiv preprint arXiv:2511.13408},
  year={2024}
}

@article{moreno2025generative,
   title={Generative flow-based warm start of the variational quantum eigensolver},
   volume={12},
   ISSN={2056-6387},
   url={http://dx.doi.org/10.1038/s41534-025-01159-x},
   DOI={10.1038/s41534-025-01159-x},
   number={1},
   journal={npj Quantum Information},
   publisher={Springer Science and Business Media LLC},
   author={Zou, Hang and Rahm, Martin and Kockum, Anton Frisk and Olsson, Simon},
   year={2025},
   month=Dec }

@inproceedings{lechner2024warm,
  title={Warm-Starting Patterns for Quantum Algorithms},
  author={Felix Truger and Johanna Barzen and Martin Beisel and Frank Leymann and and Vladimir Yussupov},
  journal={PATTERNS 2024 : The Sixteenth International Conference on Pervasive Patterns and Applications},
  year={2024}
}

@article{zhang2024warm,
  title={Warm start of variational quantum algorithms for quadratic unconstrained binary optimization problems.},
  author={Chai, Y., Jansen and K., Kühn and S. et al.},
  journal={EPJ Quantum Technol. 13, 9 (2026)},
  url={ https://doi.org/10.1140/epjqt/s40507-025-00452-0},
  year={2026}
}

@article{Eisert2010,
  title = {Colloquium: Area laws for the entanglement entropy},
  author = {Eisert, J. and Cramer, M. and Plenio, M. B.},
  journal = {Rev. Mod. Phys.},
  volume = {82},
  pages = {277--306},
  year = {2010},
  month = {Feb},
  publisher = {American Physical Society},
  doi = {10.1103/RevModPhys.82.277},
  url = {https://link.aps.org/doi/10.1103/RevModPhys.82.277}
}

@article{Cramer2006,
   title={Efficient Description of Many-Body Systems with Matrix Product Density Operators},
   volume={1},
   ISSN={2691-3399},
   url={http://dx.doi.org/10.1103/PRXQuantum.1.010304},
   DOI={10.1103/prxquantum.1.010304},
   number={1},
   journal={PRX Quantum},
   publisher={American Physical Society (APS)},
   author={Guth Jarkovský, Jiří and Molnár, András and Schuch, Norbert and Cirac, J. Ignacio},
   year={2020},
   month=Sept }

@article{Refael2009,
   title={Criticality and entanglement in random quantum systems},
   volume={42},
   ISSN={1751-8121},
   url={http://dx.doi.org/10.1088/1751-8113/42/50/504010},
   DOI={10.1088/1751-8113/42/50/504010},
   number={50},
   journal={Journal of Physics A: Mathematical and Theoretical},
   publisher={IOP Publishing},
   author={Refael, G and Moore, J E},
   year={2009},
   month=Dec, pages={504010} }

@article{Stephan2011,
   title={Logarithmic Terms in Entanglement Entropies of 2D Quantum Critical Points and Shannon Entropies of Spin Chains},
   volume={107},
   ISSN={1079-7114},
   url={http://dx.doi.org/10.1103/PhysRevLett.107.020402},
   DOI={10.1103/physrevlett.107.020402},
   number={2},
   journal={Physical Review Letters},
   publisher={American Physical Society (APS)},
   author={Zaletel, Michael P. and Bardarson, Jens H. and Moore, Joel E.},
   year={2011},
   month=July }

@article{Malz2024,
  title = {Preparation circuits for matrix product states by classical variational disentanglement},
  author = {Mansuroglu, Refik and Schuch, Norbert},
  journal = {Phys. Rev. A},
  volume = {113},
  issue = {4},
  pages = {042430},
  numpages = {15},
  year = {2026},
  month = {Apr},
  publisher = {American Physical Society},
  doi = {10.1103/7x2g-twkl},
  url = {https://link.aps.org/doi/10.1103/7x2g-twkl}
}

@article{Coopmans2024,
   title={Quantum state preparation of normal distributions using matrix product states},
   volume={10},
   ISSN={2056-6387},
   url={http://dx.doi.org/10.1038/s41534-024-00805-0},
   DOI={10.1038/s41534-024-00805-0},
   number={1},
   journal={npj Quantum Information},
   publisher={Springer Science and Business Media LLC},
   author={Iaconis, Jason and Johri, Sonika and Zhu, Elton Yechao},
   year={2024},
   month=Jan }

@article{Schuckert2024,
   title={Constant-Depth Preparation of Matrix Product States with Adaptive Quantum Circuits},
   volume={5},
   ISSN={2691-3399},
   url={http://dx.doi.org/10.1103/PRXQuantum.5.030344},
   DOI={10.1103/prxquantum.5.030344},
   number={3},
   journal={PRX Quantum},
   publisher={American Physical Society (APS)},
   author={Smith, Kevin C. and Khan, Abid and Clark, Bryan K. and Girvin, S.M. and Wei, Tzu-Chieh},
   year={2024},
   month=Sept 
   }

@article{Scarpa2020,
  title = {Projected Entangled Pair States: Fundamental Analytical and Numerical Limitations},
  author = {Scarpa, G. and Molnár, A. and Ge, Y. and García-Ripoll, J. J. and Schuch, N. and Pérez-García, D. and Iblisdir, S.},
  journal = {Phys. Rev. Lett.},
  volume = {125},
  issue = {21},
  pages = {210504},
  year = {2020},
  month = {Nov},
  publisher = {American Physical Society},
  doi = {10.1103/PhysRevLett.125.210504},
  url = {https://link.aps.org/doi/10.1103/PhysRevLett.125.210504}
}

@article{Wei2024,
  title = {Dual-Isometric Projected Entangled Pair States},
  author = {Yu, Xie-Hang and Cirac, J. Ignacio and Kos, Pavel and Styliaris, Georgios},
  journal = {Phys. Rev. Lett.},
  volume = {133},
  issue = {19},
  pages = {190401},
  numpages = {7},
  year = {2024},
  month = {Nov},
  publisher = {American Physical Society},
  doi = {10.1103/PhysRevLett.133.190401},
  url = {https://link.aps.org/doi/10.1103/PhysRevLett.133.190401}
}

@article{Jahromi2021,
   title={Simulation of three-dimensional quantum systems with projected entangled-pair states},
   volume={103},
   ISSN={2469-9969},
   url={http://dx.doi.org/10.1103/PhysRevB.103.205137},
   DOI={10.1103/physrevb.103.205137},
   number={20},
   journal={Physical Review B},
   publisher={American Physical Society (APS)},
   author={Vlaar, Patrick C. G. and Corboz, Philippe},
   year={2021},
   month=May }

@article{Verstraete2004,
  title = {Renormalization algorithms for quantum-many body systems in two and higher dimensions},
  author = {Verstraete, F. and Cirac, J. I.},
  journal = {arXiv preprint cond-mat/0407066},
  year = {2004},
  url = {https://arxiv.org/abs/cond-mat/0407066}
}

@article{Nakagawa2020,
   title={Area law of noncritical ground states in 1D long-range interacting systems},
   volume={11},
   ISSN={2041-1723},
   url={http://dx.doi.org/10.1038/s41467-020-18055-x},
   DOI={10.1038/s41467-020-18055-x},
   number={1},
   journal={Nature Communications},
   publisher={Springer Science and Business Media LLC},
   author={Kuwahara, Tomotaka and Saito, Keiji},
   year={2020},
   month=Sept }

@article{Jansen2007,
  title = {Bounds for the adiabatic approximation with applications to quantum computation},
  author = {Jansen, S. and Ruskai, M.-B. and Seiler, R.},
  journal = {J. Math. Phys.},
  volume = {48},
  pages = {102111},
  year = {2007},
  doi = {10.1063/1.2798382}
}

@article{Guo2024,
  title = {Improved gap dependence in adiabatic state preparation by adaptive schedule},
  author = {Guo, Xi and An, Dong},
  journal = {arXiv preprint arXiv:2512.10329},
  year = {2024},
  url = {https://arxiv.org/abs/2512.10329}
}

@article{Rai2024,
  title = {Spectral Gap Optimization for Enhanced Adiabatic State Preparation},
  author = {Rai, Kshiti Sneh and Chen, Jin-Fu and Emonts, Patrick and Tura, Jordi},
  journal = {arXiv preprint arXiv:2409.15433},
  year = {2024},
  url = {https://arxiv.org/abs/2409.15433}
}

@article{Wan2022,
  title = {Fast digital methods for adiabatic state preparation},
  author = {Wan, Kianna and Kim, Isaac H.},
  journal = {arXiv preprint arXiv:2004.04164},
  year = {2020},
  url = {https://arxiv.org/abs/2004.04164}
}

@article{Peruzzo2014,
  title = {A variational eigenvalue solver on a photonic quantum processor},
  author = {Peruzzo, Alberto and McClean, Jarrod and Shadbolt, Peter and Yung, Man-Hong and Zhou, Xiao-Qi and Love, Peter J. and Aspuru-Guzik, Alán and O'Brien, Jeremy L.},
  journal = {Nat. Commun.},
  volume = {5},
  pages = {4213},
  year = {2014},
  doi = {10.1038/ncomms5213}
}

@article{Marti_2025,
   title={Efficient Quantum Cooling Algorithm for Fermionic Systems},
   volume={9},
   ISSN={2521-327X},
   url={http://dx.doi.org/10.22331/q-2025-02-18-1635},
   DOI={10.22331/q-2025-02-18-1635},
   journal={Quantum},
   publisher={Verein zur Forderung des Open Access Publizierens in den Quantenwissenschaften},
   author={Marti, Lucas and Mansuroglu, Refik and Hartmann, Michael J.},
   year={2025},
   month=Feb, pages={1635} }

@misc{Preti2022,
      title={Phenomenological Theory of Variational Quantum Ground-State Preparation}, 
      author={Nikita Astrakhantsev and Guglielmo Mazzola and Ivano Tavernelli and Giuseppe Carleo},
      year={2022},
      eprint={2205.06278},
      archivePrefix={arXiv},
      primaryClass={quant-ph},
      url={https://arxiv.org/abs/2205.06278}, 
}

@article{Wang2023,
  title = {Scalable Quantum Ground State Preparation of the Heisenberg Model: A Variational Quantum Eigensolver Approach},
  author = {Wang, Jinao and Jaiswal, Rimika},
  journal = {arXiv preprint arXiv:2308.12020},
  year = {2023},
  url = {https://arxiv.org/abs/2308.12020}
}

@article{Grimsley2024,
  title = {An adaptive variational algorithm for exact molecular simulations on a quantum computer},
  author = {Grimsley, Harper R. and Economou, Sophia E. and Barnes, Edwin and Mayhall, Nicholas J.},
  journal = {Nat Commun 10, 3007},
  year = {2019},
  url = {https://doi.org/10.1038/s41467-019-10988-2}
}

@misc{Watanabe2024,
      title={Efficient ground state preparation in variational quantum eigensolver with symmetry-breaking layers}, 
      author={Chae-Yeun Park},
      year={2021},
      eprint={2106.02509},
      archivePrefix={arXiv},
      primaryClass={quant-ph},
      url={https://arxiv.org/abs/2106.02509}, 
}

@article{Shen2022,
   title={Improved Variational Quantum Eigensolver Via Quasidynamical Evolution},
   volume={19},
   ISSN={2331-7019},
   url={http://dx.doi.org/10.1103/PhysRevApplied.19.024047},
   DOI={10.1103/physrevapplied.19.024047},
   number={2},
   journal={Physical Review Applied},
   publisher={American Physical Society (APS)},
   author={Jattana, Manpreet Singh and Jin, Fengping and De Raedt, Hans and Michielsen, Kristel},
   year={2023},
   month=Feb }

@misc{Gokhale2021,
      title={Minimizing State Preparations in Variational Quantum Eigensolver by Partitioning into Commuting Families}, 
      author={Pranav Gokhale and Olivia Angiuli and Yongshan Ding and Kaiwen Gui and Teague Tomesh and Martin Suchara and Margaret Martonosi and Frederic T. Chong},
      year={2019},
      eprint={1907.13623},
      archivePrefix={arXiv},
      primaryClass={quant-ph},
      url={https://arxiv.org/abs/1907.13623}, 
}

@article{Geier2023,
   title={Few-Body Analog Quantum Simulation with Rydberg-Dressed Atoms in Optical Lattices},
   volume={4},
   ISSN={2691-3399},
   url={http://dx.doi.org/10.1103/PRXQuantum.4.020301},
   DOI={10.1103/prxquantum.4.020301},
   number={2},
   journal={PRX Quantum},
   publisher={American Physical Society (APS)},
   author={Malz, Daniel and Cirac, J. Ignacio},
   year={2023},
   month=Apr }

@article{Crescimanna2023,
   title={Quantum Control of Rydberg Atoms for Mesoscopic Quantum State and Circuit Preparation},
   volume={20},
   ISSN={2331-7019},
   url={http://dx.doi.org/10.1103/PhysRevApplied.20.034019},
   DOI={10.1103/physrevapplied.20.034019},
   number={3},
   journal={Physical Review Applied},
   publisher={American Physical Society (APS)},
   author={Crescimanna, Valerio and Taylor, Jacob and Goldberg, Aaron Z. and Heshami, Khabat},
   year={2023},
   month=Sept }

@misc{Geier2025,
      title={Shortcuts to Analog Preparation of Non-Equilibrium Quantum Lakes}, 
      author={Nik O. Gjonbalaj and Rahul Sahay and Susanne F. Yelin},
      year={2025},
      eprint={2502.03518},
      archivePrefix={arXiv},
      primaryClass={quant-ph},
      url={https://arxiv.org/abs/2502.03518}, 
}

@misc{Bhargava2026,
      title={Constant Depth Digital-Analog Counterdiabatic Quantum Computing}, 
      author={Balaganchi A. Bhargava and Shubham Kumar and Anne-Maria Visuri and Paolo A. Erdman and Enrique Solano and Narendra N. Hegade},
      year={2026},
      eprint={2601.01154},
      archivePrefix={arXiv},
      primaryClass={quant-ph},
      url={https://arxiv.org/abs/2601.01154}, 
}

@article{Volya2023,
  title = {State Preparation on Quantum Computers via Quantum Steering},
  author = {Volya, Daniel and Mishra, Prabhat},
  journal = {arXiv preprint arXiv:2302.13518},
  year = {2023},
  url = {https://arxiv.org/abs/2302.13518}
}

@article{McClean2016,
  title = {The theory of variational hybrid quantum-classical algorithms},
  author = {McClean, Jarrod R. and Romero, Jonathan and Babbush, Ryan and Aspuru-Guzik, Alán},
  journal = {New J. Phys.},
  volume = {18},
  pages = {023023},
  year = {2016},
  doi = {10.1088/1367-2630/18/2/023023}
}

@article{schaller2008adiabatic,
  title={Adiabatic preparation without quantum phase transitions},
  author={Schaller, Gernot},
  journal={Physical Review A—Atomic, Molecular, and Optical Physics},
  volume={78},
  number={3},
  pages={032328},
  year={2008},
  publisher={APS}
}

@article{kitaev1995quantum,
  author        = {Kitaev, Alexei Yu.},
  title         = {Quantum measurements and the {A}belian stabilizer problem},
  journal       = {arXiv preprint},
  year          = {1995},
  eprint        = {quant-ph/9511026},
  archivePrefix = {arXiv},
  primaryClass  = {quant-ph}
}

@article{cleve1998quantum,
  author        = {Cleve, Richard and Ekert, Artur and Macchiavello, Chiara and Mosca, Michele},
  title         = {Quantum algorithms revisited},
  journal       = {Proceedings of the Royal Society of London A},
  volume        = {454},
  number        = {1969},
  pages         = {339--354},
  year          = {1997},
  publisher     = {The Royal Society},
  eprint        = {quant-ph/9708016},
  archivePrefix = {arXiv},
  primaryClass  = {quant-ph}
}

@article{abrams1999quantum,
  author        = {Abrams, Daniel S. and Lloyd, Seth},
  title         = {Quantum algorithm providing exponential speed increase
                   for finding eigenvalues and eigenvectors},
  journal       = {Physical Review Letters},
  volume        = {83},
  number        = {24},
  pages         = {5162--5165},
  year          = {1999},
  publisher     = {American Physical Society},
  eprint        = {quant-ph/9807070},
  archivePrefix = {arXiv},
  primaryClass  = {quant-ph}
}

@article{aspuruguzik2005simulated,
  author        = {Aspuru-Guzik, Al{\'a}n and Dutoi, Anthony D. and Love, Peter J.
                   and Head-Gordon, Martin},
  title         = {Simulated quantum computation of molecular energies},
  journal       = {Science},
  volume        = {309},
  number        = {5741},
  pages         = {1704--1707},
  year          = {2005},
  publisher     = {American Association for the Advancement of Science},
  eprint        = {quant-ph/0604193},
  archivePrefix = {arXiv},
  primaryClass  = {quant-ph}
}

@article{kempe2006complexity,
  author        = {Kempe, Julia and Kitaev, Alexei and Regev, Oded},
  title         = {The complexity of the local {H}amiltonian problem},
  journal       = {SIAM Journal on Computing},
  volume        = {35},
  number        = {5},
  pages         = {1070--1097},
  year          = {2006},
  publisher     = {Society for Industrial and Applied Mathematics},
  eprint        = {quant-ph/0406180},
  archivePrefix = {arXiv},
  primaryClass  = {quant-ph}
}

@article{babbush2018lowdepth,
  title = {Low-Depth Quantum Simulation of Materials},
  author = {Babbush, Ryan and Wiebe, Nathan and McClean, Jarrod and McClain, James and Neven, Hartmut and Chan, Garnet Kin-Lic},
  journal = {Phys. Rev. X},
  volume = {8},
  issue = {1},
  pages = {011044},
  numpages = {40},
  year = {2018},
  month = {Mar},
  publisher = {American Physical Society},
  doi = {10.1103/PhysRevX.8.011044},
  url = {https://link.aps.org/doi/10.1103/PhysRevX.8.011044}
}

@article{roland2002quantum,
   title={Quantum search by local adiabatic evolution},
   volume={65},
   ISSN={1094-1622},
   url={http://dx.doi.org/10.1103/PhysRevA.65.042308},
   DOI={10.1103/physreva.65.042308},
   number={4},
   journal={Physical Review A},
   publisher={American Physical Society (APS)},
   author={Roland, Jérémie and Cerf, Nicolas J.},
   year={2002},
   month=Mar }

@article{pelofske2026vqe,
  title={VQE as Initial State Preparation for QPE on Heisenberg Spin-Glass Hamiltonians},
  author={Pelofske, Elijah and Eidenbenz, Stephan},
  journal={arXiv preprint arXiv:2606.15061},
  year={2026}
}

@article{singh2026ground,
  title={Ground-state reachability for variational quantum eigensolvers: a Rydberg-atom case study},
  author={Singh, Juhi and Kruckenhauser, Andreas and van Bijnen, Rick and Zeier, Robert},
  journal={Quantum Science and Technology},
  volume={11},
  number={3},
  pages={035033},
  year={2026},
  publisher={IOP Publishing}
}

@article{jaderberg2026learning,
  title={Learning ground state observables from quantum computing experiments},
  author={Jaderberg, Ben and Shah, Freya and Jeon, Minjun and Sahin, M Emre and Zoufal, Christa and Sharma, Kunal},
  journal={arXiv preprint arXiv:2606.15983},
  year={2026}
}

@article{bausch2022fast,
  title={Fast black-box quantum state preparation},
  author={Bausch, Johannes},
  journal={Quantum},
  volume={6},
  pages={773},
  year={2022},
  publisher={Verein zur F{\"o}rderung des Open Access Publizierens in den Quantenwissenschaften}
}

@article{patkowski2026hierarchical,
  title={Hierarchical fusion method for scalable quantum eigenstate preparation},
  author={Patkowski, Matthew and Ayyildiz, Onat and Kebri{\v{c}}, Matja{\v{z}} and Hunt, Katharine LC and Lee, Dean},
  journal={Physical Review A},
  volume={113},
  number={5},
  pages={052442},
  year={2026},
  publisher={APS}
}

\end{document}